\documentclass[aps,prx,twocolumn,showpacs,superscriptaddress]{revtex4-2}

\usepackage{graphicx}
\usepackage[colorlinks,citecolor=blue,urlcolor=blue,linkcolor=blue]{hyperref}
\usepackage{siunitx}	
\usepackage{amsmath}
\usepackage{amssymb}
\usepackage{amsthm}
\usepackage{braket}
\usepackage{booktabs}
\usepackage{footmisc}
\usepackage{mathrsfs}
\usepackage{cleveref}
\usepackage{dsfont}
\usepackage{subfigure}
\usepackage{multirow}

\newcommand{\unit}{\mathds{1}}

\newcommand{\sgn}{\mathrm{sgn}}

\newcommand{\tr}[1]{\mathrm{Tr}\left(#1\right)}
\newcommand{\trp}[2]{\mathrm{Tr}_\mathrm{#1}\left(#2\right)}
\newcommand{\abs}[1]{\left| #1 \right|} 

\newcommand{\bracket}[2]{\ensuremath{\langle#1 \vphantom{#2}| #2\vphantom{#1}\rangle}}

\newcommand{\ketbra}[2]{\ensuremath{|#1 \vphantom{#2}\rangle \langle #2\vphantom{#1}|}}

\begin{document}
\title{Wave-particle duality of many-body quantum states}

\author{Christoph Dittel}
\email{christoph.dittel@physik.uni-freiburg.de}
\affiliation{Institut f{\"u}r Experimentalphysik, Universit{\"a}t Innsbruck, Technikerstr. 25, 6020 Innsbruck, Austria}
\affiliation{Physikalisches Institut, Albert-Ludwigs-Universit{\"a}t Freiburg, Hermann-Herder-Str. 3, 79104 Freiburg, Germany}
\affiliation{EUCOR Centre for Quantum Science and Quantum Computing, Albert-Ludwigs-Universit{\"a}t Freiburg, Hermann-Herder-Str. 3, 79104 Freiburg, Germany}

\author{Gabriel Dufour}
\affiliation{Physikalisches Institut, Albert-Ludwigs-Universit{\"a}t Freiburg, Hermann-Herder-Str. 3, 79104 Freiburg, Germany}
\affiliation{EUCOR Centre for Quantum Science and Quantum Computing, Albert-Ludwigs-Universit{\"a}t Freiburg, Hermann-Herder-Str. 3, 79104 Freiburg, Germany}
\affiliation{Freiburg Institute for Advanced Studies, Albert-Ludwigs-Universit{\"a}t-Freiburg, Albertstr. 19, 79104 Freiburg, Germany}

\author{Gregor Weihs}
\affiliation{Institut f{\"u}r Experimentalphysik, Universit{\"a}t Innsbruck, Technikerstr. 25, 6020 Innsbruck, Austria}

\author{Andreas Buchleitner}
\affiliation{Physikalisches Institut, Albert-Ludwigs-Universit{\"a}t Freiburg, Hermann-Herder-Str. 3, 79104 Freiburg, Germany}
\affiliation{EUCOR Centre for Quantum Science and Quantum Computing, Albert-Ludwigs-Universit{\"a}t Freiburg, Hermann-Herder-Str. 3, 79104 Freiburg, Germany}

\date{\today}

\begin{abstract}
We formulate a general theory of wave-particle duality for many-body quantum states, which quantifies how wave- and particle-like properties balance each other. Much as in the well-understood single-particle case, which-way information -- here on the level of many-particle paths -- lends particle-character, while interference -- here due to coherent superpositions of many-particle amplitudes -- indicates wave-like properties. We analyze how many-particle which-way information, continuously tunable by the level of distinguishability of fermionic or bosonic, identical and possibly interacting particles, constrains interference contributions to many-particle observables and thus controls the quantum-to-classical transition in many-particle quantum systems. The versatility of our theoretical framework is illustrated for Hong-Ou-Mandel- and Bose-Hubbard-like exemplary settings.
\end{abstract}

\maketitle
\section{Introduction}
The coexistence of wave- and particle-like features in the behavior of quantum objects lies at the very heart of quantum theory \cite{Bohr-QM-1935} and has been contemplated since Bohr's and Einstein's early debate on the double-slit experiment \cite{Bohr-DE-1949}. According to Bohr, ``\textit{evidence obtained under different experimental conditions cannot be comprehended within a single picture, but must be regarded as complementary in the sense that only the totality of the phenomena exhausts the possible information about the objects}''~\cite{Bohr-DE-1949}. Quantitative expressions of this statement in terms of wave-particle complementarity relations \cite{Wootters-CD-1979,Greenberger-SW-1988,Mandel-CI-1991,Jaeger-TI-1995,Englert-FV-1996,Duerr-QW-2001,Bimonte-CQ-2003,Bimonte-ID-2003,Englert-WP-2008,Siddiqui-TS-2015,Bera-DQ-2015,Bagan-RB-2016,Coles-CF-2016,Qureshi-WP-2017,Bagan-DG-2018} weight which-way information against fringe visibility in 
interferometric settings. Confirmed by experimental evidence for ever larger single quantum objects \cite{Duerr-FV-1998,Mei-CD-2001,Peng-IC-2003,Hornberger-CD-2003,Hackermueller-DM-2004,Arndt-PL-2005,Schwindt-QW-1999,Jacques-DC-2008,Yuan-ED-2018}, they consolidate the fundamental status of complementarity on the single-particle level. Complementarity thus constitutes a cornerstone of our modern understanding of decoherence as the consequence of the availability of which-way information -- i.e., of the manifestation of an object's particle character -- in quantum dynamical processes. As the considered object's size increases, which-way information is easier to assess, and interference phenomena therefore become ever more fragile \cite{Brune-OP-1996}, consistently with our everyday experiences in the macroscopic world.

\begin{figure*}[t]
\centering
\includegraphics[width=\linewidth]{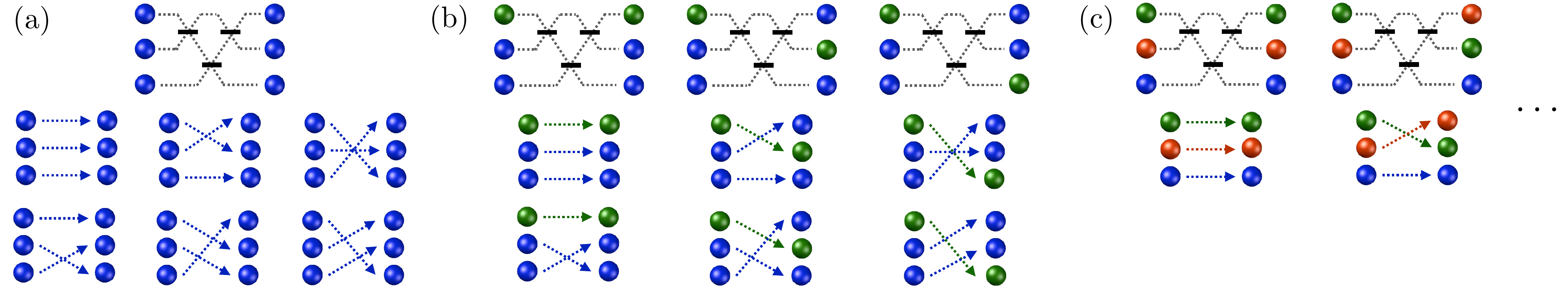}
\caption{Three-particle transition amplitudes of (a) indistinguishable 
(indicated by identical colouring), (b) partially (two different colours), and (c) fully (three different colours) distinguishable particles. In the fully indistinguishable case, six indistinguishable three-particle transition amplitudes add coherently to determine the output event's probability \cite{Tannoudji-QM-1973}. In the partially distinguishable case, the six transition amplitudes from (a) fall apart into three mutually distinct pairs of interfering two-particle amplitudes. In the fully distinguishable case, none of the -- now perfectly distinguishable -- transition amplitudes superimpose coherently any more.}
\label{fig:many-particle-paths}
\end{figure*}

However, quantum interference is not restricted to single particles, but can also arise in the evolution of ensembles of identical particles. Such interference is rooted in the inability to attribute unambiguous evolution paths to each of the ensemble's identical constituents, so that various many-particle transition amplitudes from a given input to a well-defined output state sum up coherently \cite{Hong-MS-1987,Shih-NT-1988,Tichy-ZT-2010,Mayer-CS-2011}. 
Yet, if the particles possess additional degrees of freedom (e.g., the polarization of photons, or the electronic levels of cold atoms) through which they can be (fully or partially) 
distinguished -- hereafter referred to as  \emph{internal}, in contrast to the \emph{external} degree of freedom in which interference is detected -- which-way information becomes available on the level of many particle transition amplitudes, and their interference must progressively fade away. Fig.~\ref{fig:many-particle-paths} illustrates how, for more than two particles, many-particle interference involves ever fewer particles as these
become more distinguishable. 

It is thus qualitatively clear that the interference of indistinguishable particles induces a potentially large number of interference contributions (possibly on top of single particle interference terms), and that the interference contrast in suitably chosen many-particle observables will be maximal for strictly indistinguishable particles. This suggests a many-particle version of wave-particle duality, which we here formulate in terms of quantitative complementarity relations. While the deterioration of many-particle interference phenomena by particle distinguishability is a subject of lively scientific debate \cite{Ra-NQ-2013,Shchesnovich-SC-2014,Guise-CL-2014,Shchesnovich-PI-2015,Tichy-SP-2015,Shchesnovich-TB-2015,Laibacher-PC-2015,Tamma-MC-2015,Tillmann-GM-2015,Shchesnovich-CP-2017,Shchesnovich-PD-2017,Walschaers-SB-2018,Khalid-PS-2018,Stanisic-DD-2018}, such relations have so far been unavailable. Since promising quantum information schemes as optical quantum computation \cite{Knill-SE-2001,OBrien-OQ-2007} or boson sampling \cite{Aaronson-CC-2013} exploit the interference of many non-interacting particles and have been demonstrated in small scale experiments \cite{OBrien-DA-2003,Franson-ED-2003,Gasparoni-RP-2004,Crespi-IM-2013,Broome-PB-2013,Tillmann-EB-2013,Wang-HE-2017}, we trust that quantitative complementarity relations will be valuable for benchmarking quantum computation platforms of increasing size. Another possible area of application is defined by experiments with ultracold atoms \cite{Kaufman-TW-2014,Preiss-SC-2015,Kaufman-QT-2016,Zeiher-CM-2017,Gross-QS-2017,Bergschneider-CE-2018,Kaufman-HO-2018,Preiss-HC-2019}, which additionally feature control over interactions and offer the possibility to study many-particle interference in strongly correlated quantum systems. Finally, by analogy with the discussion of wave-particle duality for single quantum objects, our results pave the way for a systematic many-particle decoherence theory, which remains to be formulated.

Our present contribution is structured as follows: In Sec.~\ref{sec:double-slit} we begin with a brief discussion of the single-particle double-slit experiment and derive two complementarity relations, one of which was not yet considered in the literature. Section~\ref{sec:pd-particles} then treats systems of many particles. Our first-quantization formalism for many partially distinguishable particles is presented in Sec.~\ref{sec:pd-particles-intro}. In Sec.~\ref{sec:reducedstates}, we examine the many-particle state's properties in terms of the involved particles' distinguishability. Measures of wave and particle character are defined in Sec.~\ref{sec:measures}, and their interdependence through wave-particle complementarity is elaborated upon in Sec.~\ref{sec:duality}. Next, in Sec.~\ref{sec:visibilities}, we consider generic many-particle interference experiments. We discuss changes in the output statistics when particles are permuted at the input in Sec.~\ref{sec:permuted}. In Sec.~\ref{sec:pdistVSindist} we derive bounds for the difference between the output statistics obtained with partially distinguishable and fully distinguishable or indistinguishable particles. We establish in Sec.~\ref{sec:pdistVSdist} visibility measures of many-particle interference signals that are fundamentally bounded by the particles' distinguishability and apply irrespective of the exact experimental scenario and of the particles' interaction strength. The versatility of these visibility measures is illustrated in Sec.~\ref{sec:examples}, where we apply our findings to the Hong-Ou-Mandel experiment and to the Bose-Hubbard model. Finally, Sec.~\ref{sec:conclusion} concludes the paper. For the sake of readability, all detailed proofs are deferred to the Appendices.

\section{Double-slit experiment}
\label{sec:double-slit}
Before we turn to the case of many particles, we start with a brief discussion of how wave-particle duality manifests in the double-slit experiment with a which-path detector acquiring partial information about the particle's path. We follow the approach of Ref.~\cite{Englert-FV-1996}, and derive two wave-particle duality relations. This establishes the basis for our considerations on wave-particle duality for many-body quantum states in the subsequent sections.

\begin{figure}[t]
\centering
\includegraphics[width=\linewidth]{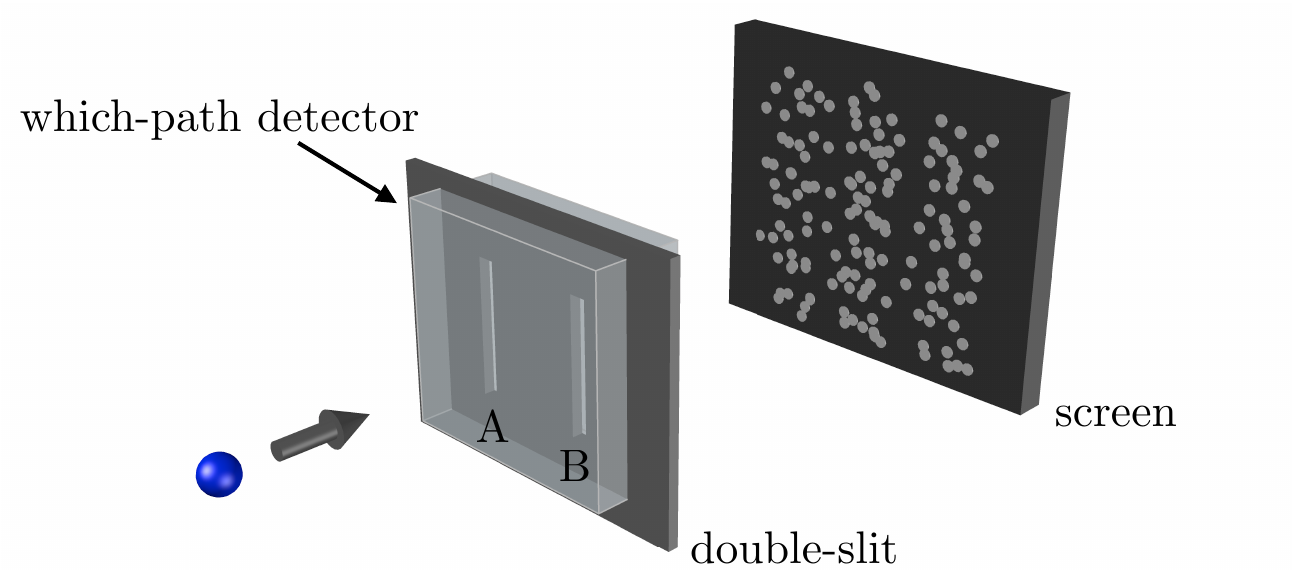}
\caption{Single-particle double-slit experiment in the presence of a which-path detector (transparent box). A single particle (blue ball) passes through a double-slit and is detected on a screen. The amount of information on the particle's path, obtained by the which-path detector, determines the visibility of the interference pattern on the screen.}
\label{fig:double-slit}
\end{figure}

Let us suppose that a single particle, initially in the pure state $\ket{P_0}$, is incident on a symmetric double-slit, with the slits labeled $\mathrm{A}$ and $\mathrm{B}$ as illustrated in Fig.~\ref{fig:double-slit}. Further, consider a which-path detector initially in a mixed state $\rho_{\mathrm{d}_0}=\sum_j q_j \ketbra{D_0^{(j)}}{D_0^{(j)}}$, i.e. in a statistical mixture of states $\ket{D_0^{(j)}}$, with probabilities $q_j\geq 0$, $\sum_j q_j=1$. Therefore, the common initial density operator of particle and detector reads
\begin{align}\label{eq:rhopdinitial}
\rho_\mathrm{pd}^\mathrm{ini}=\ket{P_0}\bra{P_0} \otimes \sum_j q_j \ketbra{D_0^{(j)}}{D_0^{(j)}}.
\end{align}

When the particle passes through the double-slit, its state becomes a balanced superposition of $\ket{P_\mathrm{A}}$ and $\ket{P_\mathrm{B}}$, with $\ket{P_\mathrm{A}}$ (resp. $\ket{P_\mathrm{B}}$) corresponding to the particle passing through slit $\mathrm{A}$ (resp. $\mathrm{B}$), and $\bracket{P_\mathrm{A}}{P_\mathrm{B}}=0$. The detector gains information about the particle's path by changing its states $\ket{D_0^{(j)}}$ to $\ket{D_\mathrm{A}^{(j)}}$ (resp. $\ket{D_\mathrm{B}^{(j)}}$) if the particle is in state $\ket{P_\mathrm{A}}$ (resp. $\ket{P_\mathrm{B}}$), where $\ket{D_\mathrm{A}^{(j)}}$ and $\ket{D_\mathrm{B}^{(j)}}$ are not necessarily orthogonal:
\begin{align*}
\ket{P_0} \otimes \ket{D_0^{(j)}} \mapsto \frac{1}{\sqrt{2}}\left(\ket{P_\mathrm{A}}\otimes \ket{D_\mathrm{A}^{(j)}}+\ket{P_\mathrm{B}}\otimes \ket{D_\mathrm{B}^{(j)}}  \right).
\end{align*}
Thereby, particle and detector become entangled, and their common state~\eqref{eq:rhopdinitial} then reads
\begin{align}\label{eq:rhoPD}
\rho_\mathrm{pd}=\frac{1}{2} \sum_{J,K\in\{\mathrm{A},\mathrm{B}\}} \ketbra{P_J}{P_K} \otimes \sum_j q_j \ketbra{D_J^{(j)}}{D_K^{(j)}}.
\end{align}

The reduced state of the particle is obtained by taking the partial trace over the detector subsystem,
\begin{align}\label{eq:rhoP}
\rho_\mathrm{p}=\trp{d}{\rho_\mathrm{pd}}= \sum_{J,K\in\{\mathrm{A},\mathrm{B}\}} [\rho_\mathrm{p}]_{J,K} \ketbra{P_J}{P_K},
\end{align}
with 
\begin{align}\label{eq:rhoPelement}
[\rho_\mathrm{p}]_{J,K}=1/2 \sum_j q_j \bracket{D_K^{(j)}}{D_J^{(j)}}.
\end{align}
Consequently, in the basis $\{\ket{P_\mathrm{A}},\ket{P_\mathrm{B}}\}$ of the particle, the off-diagonal element $[\rho_\mathrm{p}]_{\mathrm{A},\mathrm{B}}$ of the particle's state is governed by the overlaps $\bracket{D_\mathrm{B}^{(j)}}{D_\mathrm{A}^{(j)}}$ between the different detector states. The magnitude of this off-diagonal element also quantifies the fringe visibility $\mathcal{V}$ of the interference pattern accumulated upon repeated particle detection on the screen,
\begin{align}\label{eq:V} 
\mathcal{V}=\sum_{\substack{J,K\in\{\mathrm{A},\mathrm{B}\}\\ J\neq K}} \big| \bra{P_J}\rho_\mathrm{p}\ket{P_K}\big| =\abs{\sum_j q_j \bracket{D_\mathrm{B}^{(j)}}{D_\mathrm{A}^{(j)}}},
\end{align}
which we can thus interpret as a measure of the wave character, with the range $0\leq \mathcal{V} \leq 1$. Moreover, the visibility, and thus the wave character, quantifies the entanglement between particle and detector, as apparent by its relation to the purity of the reduced state of the particle, 
\begin{align}\label{eq:SPP}
\mathcal{V}=\sqrt{2\ \tr{\rho_\mathrm{p}^2}-1}.
\end{align}

If, instead, we trace out the particle in Eq.~\eqref{eq:rhoPD}, we obtain the reduced detector state
\begin{align}\label{eq:rhoD}
\rho_\mathrm{d}=\trp{p}{\rho_\mathrm{pd}}=\frac{1}{2} \sum_{J\in\{\mathrm{A},\mathrm{B}\}}  \rho_\mathrm{d}^{J},
\end{align}
with $\rho_\mathrm{d}^{J}=\sum_j q_j \ketbra{D_J^{(j)}}{D_J^{(j)}}$. The detector thus ends up in a balanced mixture of $\rho_\mathrm{d}^{\mathrm{A}}$ and $\rho_\mathrm{d}^{\mathrm{B}}$, which correspond to detection of the particle in slit $\mathrm{A}$ or $\mathrm{B}$, respectively. In turn, the ability to discriminate these two states via a general measurement on the detector state is related to the possibility of tracking the particle and provides a measure of the particle character.

We therefore need to compare two states $\rho$ and $\sigma$, what we accomplish by use of either their trace distance  $D(\rho,\sigma)=\tr{|\rho-\sigma|}/2$, or their (square root quantum) fidelity $F(\rho,\sigma)=\tr{\sqrt{\sqrt{\rho}\sigma\sqrt{\rho}}}$ \cite{Nielsen-QC-2011}. Here $|M|=\sqrt{M^\dagger M}$, and $\sqrt{\cdot}$ denotes the positive square root of a positive semidefinite matrix.
Both quantities take values between zero and one and they obey the Fuchs-van de Graaf inequalities \cite{Fuchs-CD-1999}
\begin{align}\label{eq:FvdG}
1-F(\rho,\sigma) \leq D(\rho,\sigma)\leq \sqrt{1-F^2(\rho,\sigma)}.
\end{align}
If $\rho$ or $\sigma$ is pure, the lower bound can be made tighter, $1-F^2(\rho,\sigma) \leq D(\rho,\sigma)$, and if both $\rho$ and $\sigma$ are pure, the upper bound saturates \cite{Nielsen-QC-2011}.

With this in mind, we define two measures:
\begin{align}\label{eq:DT} 
\mathcal{P}_\mathrm{t}=D(\rho_\mathrm{d}^{\mathrm{A}},\rho_\mathrm{d}^{\mathrm{B}}),
\end{align}
and 
\begin{align}\label{eq:DF} 
\mathcal{P}_\mathrm{f}=\sqrt{1-F^2(\rho_\mathrm{d}^{\mathrm{A}},\rho_\mathrm{d}^{\mathrm{B}})}.
\end{align}
These measures quantify the ability to discriminate the which-path detector states $\rho_\mathrm{d}^{\mathrm{A}}$ and $\rho_\mathrm{d}^{\mathrm{B}}$, and we identify this ability as a particle character. It immediately follows from these definitions and Eq.~\eqref{eq:FvdG} that
\begin{align}\label{eq:DDinequDouble}
\mathcal{P}_\mathrm{t} \leq \mathcal{P}_\mathrm{f},
\end{align}
with equality for pure detector states. 

Note that, in the literature \cite{Jaeger-TI-1995,Englert-FV-1996,Bimonte-CQ-2003,Bimonte-ID-2003,Siddiqui-TS-2015,Bera-DQ-2015,Bagan-RB-2016,Qureshi-WP-2017}, measures of the particle character are sometimes denoted by $\mathcal{D}$. Here we refer to measures of the particle character by $\mathcal{P}$, and reserve $\mathcal{D}$ for many-particle distinguishability measures defined further below. 

 While we elaborate on quantum state discrimination in section~\ref{sec:particlecharacter}, let us stress already here that the particle measure $\mathcal{P}_\mathrm{t}$ is related to $P_\mathrm{AQSD}$, the maximal success probability for an \emph{ambiguous} quantum state discrimination \cite{Helstrom-QD-1976,Sacchi-OD-2005,Spehner-QC-2014} between $\rho_\mathrm{d}^{\mathrm{A}}$ and $\rho_\mathrm{d}^{\mathrm{B}}$, by $P_\mathrm{AQSD}=(1+\mathcal{P}_\mathrm{t})/2$. The second particle measure, $\mathcal{P}_\mathrm{f}$, can likewise be motivated by state discrimination, since $P_\mathrm{UQSD}$, the maximal success probability for an \emph{unambiguous} quantum state discrimination \cite{Rudolph-UD-2003,Feng-UD-2004} of $\rho_\mathrm{d}^{\mathrm{A}}$ and $\rho_\mathrm{d}^{\mathrm{B}}$, obeys $P_\mathrm{UQSD}\leq 1-F(\rho_\mathrm{d}^{\mathrm{A}},\rho_\mathrm{d}^{\mathrm{B}}) \leq \mathcal{P}_\mathrm{f}$. 

As we prove in Appendix~\ref{appsec:doubleslit}, both measures $\mathcal{P}_\mathrm{t}$ and $\mathcal{P}_\mathrm{f}$ of the particle character, together with the measure $\mathcal{V}$ of the wave character, can be combined to obey wave-particle duality relations. Specifically, we find
\begin{align}\label{eq:Englert2}
\mathcal{P}_\mathrm{t}^2+\mathcal{V}^2 \leq \mathcal{P}_\mathrm{f}^2+\mathcal{V}^2 \leq 1,
\end{align}
with both inequalities saturating for pure which-path detector states. The relation $\mathcal{P}_\mathrm{t}^2+\mathcal{V}^2 \leq 1$ was proven in Refs.~\cite{Jaeger-TI-1995,Englert-FV-1996}. However, with the second inequality in Eq.~\eqref{eq:Englert2} we identify a tighter wave-particle duality relation for mixed detector states. Note that, since all above measures are normalized, i.e. $0\leq \mathcal{V},\mathcal{P}_\mathrm{t},\mathcal{P}_\mathrm{f} \leq 1$, wave and particle character are mutually exclusive, i.e. the single quantum objects under consideration cannot fully display both properties simultaneously. We stress that we refer to relations of the form~\eqref{eq:Englert2} as wave-particle \emph{duality relations} if the inequality saturates for pure states, since this case is fully characterized by quantifiers of precisely two complementary properties. Otherwise, the two involved measures do not necessarily account for the totality of all observable phenomena, and, on the basis of Bohr's notion~\cite{Bohr-DE-1949} [see the Introduction], we refer to them as \emph{complementarity relations}.

\section{Many partially distinguishable particles}
\label{sec:pd-particles}
We now proceed to our original findings for the case of many particles. We first provide a general description of partially distinguishable particles in first quantization, where we distinguish between external and internal degrees of freedom. As we elaborate upon in Sec.~\ref{sec:visibilities}, the former evolve dynamically and are resolved by the detection apparatus, while the latter are fixed during the particles' evolution and remain unresolved by the detection but can serve as \emph{labels} to distinguish the particles. We then inspect the reduced many-particle density operators obtained by tracing over the internal or external degrees of freedom and distill wave-particle duality of many-body quantum states in the same spirit as for the single-particle double-slit experiment. 

\subsection{Partially distinguishable particles in first quantization}
\label{sec:pd-particles-intro}
The first-quantization formalism developed in the present section is equivalent to the usual second-quantization approach but highlights the interdependencies between internal and external degrees of freedom, which are at the heart of our understanding of particle distinguishability. We aim at describing $N$ identical bosons or fermions which are 
prepared in an arbitrary state of their internal ($\mathrm{I}$) degrees of freedom, and expanded over $n$ mutually orthogonal external ($\mathrm{E}$) states (or modes). The single-particle Hilbert space $\mathcal{H}=\mathcal{H}_\mathrm{E}\otimes \mathcal{H}_\mathrm{I}$ is composed of the $n$-dimensional external Hilbert space $\mathcal{H}_\mathrm{E}$ spanned by the orthonormal basis $\{\ket{1},\dots,\ket{n}\}$, tensored with the $m$-dimensional internal Hilbert space $\mathcal{H}_\mathrm{I}$ spanned by the orthonormal basis $\{\ket{i_1},\dots,\ket{i_m}\}$. Note that, while we choose to work with finite-dimensional Hilbert spaces for simplicity, our formalism allows a straightforward extension to include continuous degrees of freedom. For $N$ identical particles, the basis states of $\mathcal{H}^{\otimes N}=\mathcal{H}_\mathrm{E}^{\otimes N}\otimes \mathcal{H}_\mathrm{I}^{\otimes N}$ are then given as $N$-fold tensor products of single particle basis states. An orthonormal basis of the $n^N$-dimensional external Hilbert space $\mathcal{H}_\mathrm{E}^{\otimes N}$ is therefore composed of the states
\begin{align}\label{eq:basisStateE}
\ket{\vec{\mathcal{E}}}=\ket{\mathcal{E}_1}\otimes \dots \otimes \ket{\mathcal{E}_N},
\end{align}
with $\mathcal{E}_j\in\{1,\dots,n\}$. In the literature, the $N$-tuple $\vec{\mathcal{E}}$ is commonly called \emph{mode assignment list} \cite{Tichy-ZT-2010,Tichy-MP-2012,Tichy-II-2014,Dittel-TD1-2018,Dittel-TD2-2018}. Analogously, an orthonormal basis of the $m^N$-dimensional $N$-particle internal Hilbert space $\mathcal{H}_\mathrm{I}^{\otimes N}$ is given by states
\begin{align}\label{eq:Ibasis}
\ket{\vec{\mathcal{I}}}=\ket{\mathcal{I}_1}\otimes \dots \otimes \ket{\mathcal{I}_N},
\end{align}
where $\mathcal{I}_j\in\{i_1,\dots,i_m\}$. 
Note that in Eqs.~\eqref{eq:basisStateE} and~\eqref{eq:Ibasis} each particle is implicitly given a label corresponding to its position in the tensor product. This sort of labeling is characteristic of the first quantization formalism and is unphysical for identical particles. Therefore, we eliminate it by (anti)symmetrization later on.

With an orthonormal basis of $\mathcal{H}^{\otimes N}$ at hand, we can now describe partially distinguishable particles with an arbitrary internal state and a fixed particle occupation in the external modes (the latter is a natural assumption, inspired by a typical experimental scenario, e.g., in photonic circuitry \cite{Tillmann-EB-2013,Spagnolo-EV-2014,Carolan-UL-2015,Loredo-BS-2017,Wang-HE-2017}). This distribution is specified by the \emph{mode occupation list} \cite{Tichy-ZT-2010,Tichy-MP-2012,Tichy-II-2014,Dittel-TD1-2018,Dittel-TD2-2018} $\vec{R}=(R_1,\dots,R_n)$, where $R_j$ is the number of particles in mode $j$. Since several mode assignment lists $\vec{\mathcal{E}}$ correspond to a given occupation $\vec{R}$, we single out the mode assignment list $\vec{E}\equiv\vec{E}(\vec{R})$ with components listed in non-decreasing order, $E_1\leq E_2\leq\dots\leq E_N$. Other external basis states $\ket{\vec{\mathcal{E}}}$ corresponding to the same mode occupation $\vec{R}$ are then obtained by permutation of the factors of $\ket{\vec{E}}$.
The external state of the particles can therefore be written in terms of the states
\begin{align}\label{eq:Emu}
\ket{\vec{E}_\pi}=\ket{E_{\pi(1)}}\otimes \dots \otimes \ket{E_{\pi(N)}},
\end{align}
for $\pi$ a permutation in the symmetric group $\mathrm{S}_N$. In short, for each mode occupation $\vec{R}$, there exists a unique basis state $\ket{\vec{E}}$ with elements in non-decreasing order, from which we obtain all basis states $\ket{\vec{\mathcal{E}}}$ associated with $\vec{R}$ by permuting the factors of $\ket{\vec{E}}$.

Regarding the internal degrees of freedom, we impose no restriction on the $N$-particle state, which we write as a general superposition of all internal basis states~\eqref{eq:Ibasis},
\begin{align}\label{eq:Omega}
\ket{\Omega^{(j)}}=\sum_{\vec{\mathcal{I}}} C_{\vec{\mathcal{I}}}^{(j)} \ket{\vec{\mathcal{I}}}.
\end{align}
This allows us to consider the effects of correlations and mixedness in the internal state (with the index $j$ used below to label different states in a statistical mixture) on many-particle interference in the external degree of freedom. The coefficients $C_{\vec{\mathcal{I}}}^{(j)}$ ultimately determine the distinguishability of the particles \cite{Shchesnovich-TB-2015,Shchesnovich-UG-2016,Shchesnovich-CP-2017}. The bosonic (fermionic) state of particles in the external configuration $\vec{R}$ with internal state $\ket{\Omega^{(j)}}$ is then obtained by (anti)symmetrization, i.e. by forming the coherent sum over all permutations (disregarding normalization, for the moment)
\begin{align} \label{eq:PsiSymet}
\ket{\Psi^{(j)}}\propto \sum_{\pi\in\mathrm{S}_N} (-1)^\pi_\mathrm{B(F)} \ket{\vec{E}_\pi} \otimes\left( \sum_{\vec{\mathcal{I}}} C_{\vec{\mathcal{I}}}^{(j)} \ket{\vec{\mathcal{I}}_\pi}\right) ,
\end{align}
where we have introduced the permuted internal basis states
\begin{align}
\label{eq:Imu}
\ket{\vec{\mathcal{I}}_\pi}=\ket{\mathcal{I}_{\pi(1)}}\otimes \dots \otimes \ket{\mathcal{I}_{\pi(N)}}.
\end{align}
The sign factor is given by $(-1)^\pi_\mathrm{B} =1$ for bosons and $(-1)^\pi_\mathrm{F}=\sgn(\pi)$ for fermions, respectively. As apparent from Eq.~\eqref{eq:PsiSymet}, the (anti)symmetrization of $\ket{\Psi^{(j)}}$ results in the entanglement of the particles' external and internal degrees of freedom, with a strength which depends on the particles' internal state. We will elaborate upon this observation further down.

In the case of multiple occupations, i.e.~if there is a mode $i$ such that $R_i\geq2$, distinct permutations $\pi\neq \pi'$ can lead to the same state $\ket{\vec{E}_\pi} =\ket{\vec{E}_{\pi'}}$. This defines an equivalence relation $\pi \sim \pi'$, and we constitute a set $\Sigma\equiv\Sigma(\vec R)$ by choosing one representative in each of the $R=N!/(\prod_{j=1}^n R_j!)$ equivalence classes. Any permutation $\pi\in\mathrm{S}_N$ can then be uniquely decomposed as $\pi=\xi \mu$, with $\mu\in\Sigma$ and  $\xi  \in \mathrm{S}_{\vec{R}}$, where  $\mathrm{S}_{\vec{R}}=\mathrm{S}_{R_1} \otimes \dots \otimes \mathrm{S}_{R_n}$ denotes the subgroup of $\mathrm{S}_N$ which leaves $\ket{\vec{E}}$ invariant. In group theoretical terms, $\Sigma$ is a transversal of the set of right cosets of $\mathrm{S}_{\vec{R}}$ in $\mathrm{S}_N$, also called right transversal of $\mathrm{S}_{\vec{R}}$ in $\mathrm{S}_N$ \cite{Baumslag-SO-1968}. Since each permutation $\mu \in \Sigma$ corresponds to one of the $R$ inequivalent ways of  ordering the particles, we refer to $\mu$ as a \emph{particle labeling}.
The normalized state $\ket{\Psi^{(j)}}$ can thus be written as a sum over particle labelings
\begin{align}\label{eq:PsiR}
\ket{\Psi^{(j)}}&= \frac{1}{\sqrt{R}}\sum_{\mu\in\Sigma} (-1)^\mu_\mathrm{B(F)} \ket{\vec{E}_\mu} \otimes  \ket{\Omega_\mu^{(j)}},
\end{align}
with
\begin{align}\label{eq:Omegamu}
\ket{\Omega_\mu^{(j)}}&\propto 
  \sum_{\vec{\mathcal{I}}} \left( \sum_{\xi\in\mathrm{S}_{\vec{R}} }(-1)^\xi_\mathrm{B(F)} C_{\vec{\mathcal{I}}_\xi}^{(j)}\right)  \ket{\vec{\mathcal{I}}_\mu}.
\end{align}
From Eq.~\eqref{eq:Omegamu}, one sees that the coefficients $C_{\vec{\mathcal{I}}}^{(j)}$ must be (anti)symmetrized over permutations of bosons (fermions) belonging to the same external mode, 
\begin{align*}
C_{\vec{\mathcal{I}}_\xi}^{(j)}=(-1)_\mathrm{B(F)}^\xi C_{\vec{\mathcal{I}}}^{(j)}
\end{align*}
for all $\xi\in\mathrm{S}_{\vec{R}}$ (note that $\vec{\mathcal{I}}_\xi=(\mathcal{I}_{\xi(1)},\dots,\mathcal{I}_{\xi(N)})$). For fermions, this enforces Pauli's exclusion principle \cite{Pauli-ZA-1925}: fermions in the same mode must be in orthogonal internal states. Without loss of generality, we choose the $C_{\vec{\mathcal{I}}}^{(j)}$ to already satisfy this symmetry and obey $\sum_{\vec{\mathcal{I}}} |C_{\vec{\mathcal{I}}}^{(j)}|^2=1$, such that the normalized internal states in Eq.~\eqref{eq:PsiR} read
\begin{align}\label{eq:Omegamusym}
 \ket{\Omega_\mu^{(j)}}&=
 \sum_{\vec{\mathcal{I}}}  C_{\vec{\mathcal{I}}}^{(j)} \ket{\vec{\mathcal{I}}_\mu}.
\end{align}
Finally, states with mixed internal degrees of freedom can be expressed as
\begin{align}\label{eq:rho0}
\rho=\sum_j q_j \ket{\Psi^{(j)}}\bra{\Psi^{(j)}},
\end{align}
where $\ket{\Psi^{(j)}}$ from Eq.~\eqref{eq:PsiR} appears with probability $q_j$, and $\sum_j q_j=1$.

The state of $N$ fully indistinguishable identical particles is obtained by assigning the same  internal state to each particle, e.g. $\ket{\Omega}=\ket{i_1}^{\otimes N}$. Fully distinguishable identical particles are obtained when all particles are in mutually orthogonal internal states, such that they can be identified unambiguously, e.g. if $\ket{\Omega}=\ket{i_1}\otimes\ket{i_2}\otimes\dots\otimes\ket{i_N}$. 

\begin{figure}[t]
\centering
\includegraphics[width=\linewidth]{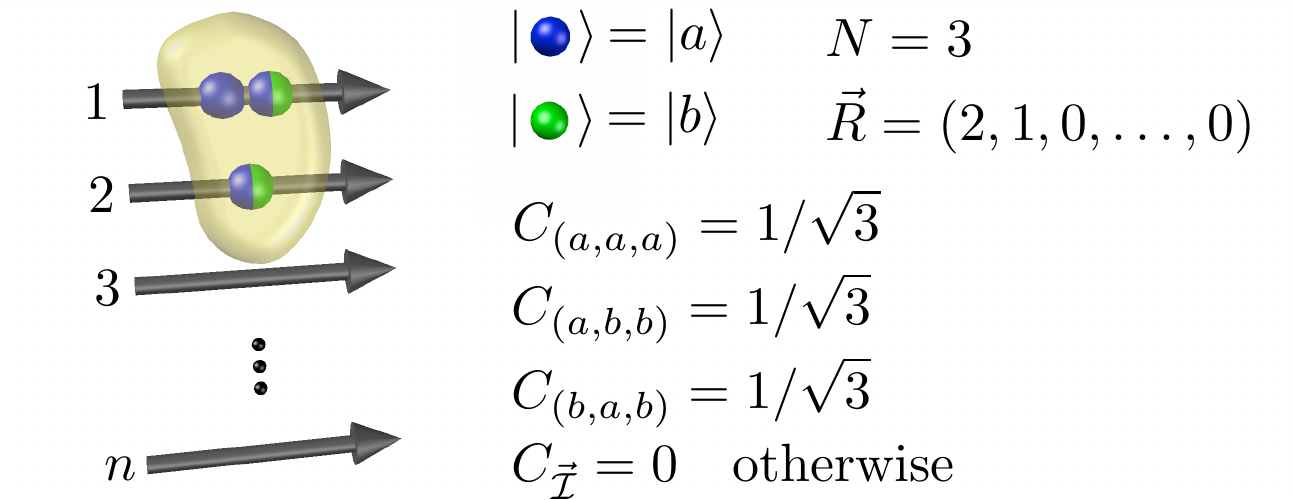}
\caption{Example of a state of $N=3$ partially distinguishable particles with mode occupation list $\vec{R}=(2,1,0,\dots,0)$ and, correspondingly, mode assignment list $\vec{E}=(1,1,2)$. External states are depicted by black arrows, particles by colored balls, and internal states by the balls' coloring, with the yellow envelope illustrating correlations between the particles. Particle distinguishability is determined by the listed coefficients $C_{\vec{\mathcal{I}}}$.}
\label{fig:Example}
\end{figure}

Before we continue, let us consider a brief example, illustrated in Fig.~\ref{fig:Example}. We consider $N=3$ bosons in $n$ modes with two particles in mode $1$ and one particle in mode $2$, such that $\vec{R}=(2,1,0\dots,0)$, $R=N!/(\prod_{j=1}^n R_j!)=3$, and $\vec{E}=(1,1,2)$. With $\epsilon$ the identity permutation and permutations given in cycle notation, we have $\mathrm{S}_{\vec{R}}=\{\epsilon, (12)\}$, and find $\Sigma=\{\epsilon, (13), (23)\}$. Note that for $\vec{R}=(2,1,0,\dots,0)$ and $\pi\in\mathrm{S}_3$, there are three distinct right cosets $\mathrm{S}_{\vec{R}} \pi$ of $\mathrm{S}_{\vec{R}}$ in $\mathrm{S}_3$, $\mathrm{S}_{\vec{R}} \epsilon=\mathrm{S}_{\vec{R}} (12)=\{\epsilon,(12)\}$, $\mathrm{S}_{\vec{R}} (13)=\mathrm{S}_{\vec{R}} (132)=\{(13),(132)\}$, and $\mathrm{S}_{\vec{R}} (23)=\mathrm{S}_{\vec{R}} (123)=\{(23),(123)\}$. The right transversal $\Sigma$ is then obtained by choosing one element of each distinct right coset, e.g. $\Sigma=\{\epsilon, (13), (23)\}$. With the help of $\Sigma$ we can extract the subset of external basis states needed in Eq.~\eqref{eq:PsiR}, $\{\ket{\vec{E}_\mu}\}_{\mu\in\Sigma}=\{ \ket{1,1,2},\ket{2,1,1},\ket{1,2,1} \}$.
In this example, we consider pure internal states, such that the sum in Eq.~\eqref{eq:rho0} consists of just one term and the index $j$ is dropped. Moreover, the single-particle internal Hilbert space is assumed to be of dimension $m=2$ and spanned by the basis $\{\ket{a},\ket{b}\}$. We consider the correlated internal state defined by the choice of coefficients $C_{(a,a,a)}=C_{(a,b,b)}=C_{(b,a,b)}=1/\sqrt{3}$, and $C_{\vec{\mathcal{I}}}=0$ otherwise. 
These coefficients satisfy the required symmetry $C_{\vec{\mathcal{I}}_\xi}=C_{\vec{\mathcal{I}}}$ for all $\xi\in \mathrm{S}_{\vec{R}}=\mathrm{S}_2\otimes\mathrm{S}_1\otimes\mathrm{S}_0\otimes\dots=\{\epsilon, (12)\}$, where permutations $\xi\in\mathrm{S}_{\vec{R}}$ only permute particles in the same mode. The internal state~\eqref{eq:Omega} then reads
\begin{align}\label{eq:Omegaexample}
\ket{\Omega}=\frac{1}{\sqrt{3}}\left( \ket{a,a,a}+\ket{a,b,b}+\ket{b,a,b}\right).
\end{align}
The symmetrized pure state~\eqref{eq:PsiR} is obtained by combining the external states $\ket{\vec{E}_\mu}$ and the internal states $\ket{\Omega_\mu}$ for all particle labelings $\mu\in\Sigma$,
\begin{align*}
\ket{\Psi}=\frac{1}{3}\big[ &\ket{1,1,2}\otimes\left( \ket{a,a,a}+\ket{a,b,b}+\ket{b,a,b}\right)\\
+&\ket{2,1,1}\otimes\left( \ket{a,a,a}+\ket{b,b,a}+\ket{b,a,b}\right)\\
+&\ket{1,2,1}\otimes\left( \ket{a,a,a}+\ket{a,b,b}+\ket{b,b,a}\right)\big],
\end{align*}
where the summands in the first, second and third row correspond to $\mu=\epsilon$, $\mu=(13)$ and $\mu=(23)$, respectively. One can easily verify that this state is symmetric under the exchange of any two particles, as required for a state of many bosons.

\subsection{The reduced external and internal states}\label{sec:reducedstates}
We now consider the reduced external and internal states of systems of partially distinguishable particles, to reveal the connection to the single-particle case described in section~\ref{sec:double-slit}. First, let us rewrite the state~\eqref{eq:rho0} in the same form as Eq.~\eqref{eq:rhoPD}, with the help of Eq.~\eqref{eq:PsiR}:
\begin{align}\label{eq:rho}
\rho=\frac{1}{R} \sum_{\mu,\nu\in\Sigma} (-1)^{\mu\nu}_\mathrm{B(F)} \ket{\vec{E}_\mu}\bra{\vec{E}_\nu} \otimes \sum_j q_j \ket{\Omega_\mu^{(j)}}\bra{\Omega_\nu^{(j)}}.
\end{align}
In general, the (anti)symmetrized state~\eqref{eq:rho} shows entanglement between external and internal degrees of freedom. As we show below, this entanglement can be used to characterize the distinguishability of the particles \cite{Stanisic-DD-2018}, and to quantify the many-particle wave character.

We first consider the \emph{reduced external state} obtained by tracing out the internal state-space in Eq.~\eqref{eq:rho},
\begin{align}\label{eq:rhoE}
 \rho_\mathrm{E}=\trp{I}{\rho}= \sum_{\mu,\nu\in\Sigma} [\rho_\mathrm{E}]_{\mu,\nu} \ket{\vec{E}_\mu}\bra{\vec{E}_\nu} ,
\end{align}
with elements 
\begin{align}\label{eq:rhoEoff-diag}
[\rho_\mathrm{E}]_{\mu,\nu}=(-1)^{\mu\nu}_\mathrm{B(F)}\frac{1}{R}\sum_j q_j \bracket{\Omega_\nu^{(j)}}{\Omega_\mu^{(j)}}
\end{align}
[compare with Eqs.~\eqref{eq:rhoP} and~\eqref{eq:rhoPelement}]. In the basis $\{\ket{\vec{\mathcal{E}}}\}_{\vec{\mathcal{E}}}$, the off-diagonal elements of $\rho_\mathrm{E}$ are thus determined by particle distinguishability via the overlaps $\bracket{\Omega_\nu^{(j)}}{\Omega_\mu^{(j)}}$ for different particle labelings $\mu\neq\nu$. Note that in Refs.~\cite{Shchesnovich-PI-2015,Shchesnovich-TB-2015,Shchesnovich-CP-2017}, particle distinguishability is described by the so called $J$-matrix which, in our formalism, has elements \unexpanded{$J_{\nu,\mu}=R [\rho_\mathrm{E}]_{\mu,\nu}=\sum_j q_j \bracket{\Omega_\nu^{(j)}}{\Omega_\mu^{(j)}}$}. On the other hand, for the description of particles in an internal product state, Ref.~\cite{Tichy-SP-2015} introduced the distinguishability matrix $\mathcal{S}$. In this case, we have \unexpanded{$\bracket{\Omega_\nu}{ \Omega_\mu}=\prod_{i=1}^N \mathcal{S}_{\nu (i),\mu (i)}$}  [see also Ref.~\cite{Chin-PD-2018}].

In the case of distinguishable particles $(\mathrm{D})$, each particle is in a distinct orthogonal internal state and these overlaps vanish. As a result, the reduced external state is maximally mixed,
\begin{align}\label{eq:rhoEdist}
\rho_\mathrm{E}^{\mathrm{D}}=\frac{1}{R} \sum_{\mu\in\Sigma} \ket{\vec{E}_\mu}\bra{\vec{E}_\mu}.
\end{align}
On the other hand, perfectly indistinguishable bosons (fermions) share the same pure internal state, such that $\bracket{\Omega_\nu^{(j)}}{\Omega_\mu^{(j)}}=1$, for all $j$ and $\mu,\nu\in\Sigma$.
Therefore, the external state is pure and given by
\begin{align}\label{eq:rhoEBF}
\rho_\mathrm{E}^{\mathrm{B}(\mathrm{F})}=\ket{\psi_{\mathrm{B}(\mathrm{F})}}\bra{\psi_{\mathrm{B}(\mathrm{F})}},
\end{align}
with
\begin{align}\label{eq:rhoEBFpure}
\ket{\psi_{\mathrm{B}(\mathrm{F})}}=\frac{1}{\sqrt{R}}\sum_{\mu\in\Sigma} (-1)^{\mu}_\mathrm{B(F)} \ket{\vec{E}_\mu}.
\end{align}
The coherences (off-diagonal elements) of the reduced external state~\eqref{eq:rhoE} thus reflect the indistinguishability of the particles. As we show in Eq.~\eqref{eq:pj-decomp} in Sec.~\ref{sec:visibilities} below, these coherences are at the origin of many-particle interference in the external degrees of freedom and are therefore constitutive of the particles' wave character.
On a related note, the purity of $\rho_\mathrm{E}$, which, by virtue of Eqs.~\eqref{eq:rhoEdist} and~\eqref{eq:rhoEBF}, quantifies the separability of internal and external degrees of freedom, is, in turn, a marker of indistinguishability. In the case of indistinguishable particles, the particles' internal and external degrees of freedom are uncorrelated and $\rho_\mathrm{E}$ appears pure. On the other hand, for fully distinguishable particles, with each particle in a distinct orthogonal internal state, the particles' internal and external degrees of freedom are maximally correlated, and the reduced state $\rho_\mathrm{E}^\mathrm{D}$ is maximally mixed on its support. We pursue this direction further in the next section.

Let us now consider the \emph{reduced internal state} by tracing over the external state-space in Eq.~\eqref{eq:rho}:
\begin{align}\label{eq:rhoI}
\rho_\mathrm{I}=\trp{E}{\rho}=\frac{1}{R} \sum_{\mu\in\Sigma} \rho_\mathrm{I}^{\mu}.
\end{align}
The result is a balanced mixture of the internal states 
\begin{align}\label{eq:rhoImu}
\rho_\mathrm{I}^{\mu}=\sum_j q_j \ket{\Omega^{(j)}_\mu}\bra{\Omega^{(j)}_\mu},
\end{align} 
which correspond to different particle labelings $\mu$ [note the close analogy with Eq.~\eqref{eq:rhoD}]. 
In the case of indistinguishable particles, the internal states $\rho_\mathrm{I}^{\mu}$ are equal for all particle labelings $\mu\in\Sigma$ and cannot be discriminated. In contrast, for distinguishable particles and pure internal states, such that the sum in~\eqref{eq:rhoImu} reduces to a single term, the states $\rho_\mathrm{I}^{\mu}$ can be discriminated with certainty, i.e. $\rho_\mathrm{I}^{\mu}$ and $ \rho_\mathrm{I}^{\nu}$ have support on orthogonal subspaces for $\mu\neq\nu$. Therefore, different particle labelings can be told apart from each other. The ability to discriminate different labelings thus constitutes a particle-like property of the many-body state. Note, however, that, for mixed internal states, even if every term in the mixture corresponds to fully distinguishable particles, perfect discrimination of the labellings might not be possible.

\subsection{Measures for wave and particle character}
\label{sec:measures}
In the previous section, we identified the magnitude of the coherences of the reduced external state~\eqref{eq:rhoE} with the many-particle wave character, and the ability to discriminate the internal states~\eqref{eq:rhoImu} with the constituents' particle character. Based on these observations, we now define normalized measures that quantify these attributes.

\subsubsection{Wave character}
A first measure for the wave character of a many-particle state $\rho$ is given by the \emph{normalized coherence} of $\rho_\mathrm{E}$,
\begin{align}\label{eq:Cmeasure}
\mathcal{W}_\mathrm{C}=\frac{1}{R-1} \sum_{\substack{\mu,\nu \in\Sigma\\ \mu\neq\nu}} \abs{\bra{\vec{E}_\mu}\rho_\mathrm{E}\ket{\vec{E}_\nu}}. 
\end{align}
This is simply the sum of absolute values of off-diagonal elements of $\rho_\mathrm{E}$, normalized such that $0\leq\mathcal{W}_\mathrm{C}\leq 1$, with $\mathcal{W}_\mathrm{C}=0$ for distinguishable and $\mathcal{W}_\mathrm{C}=1$ for indistinguishable particles [compare to Eq.~\eqref{eq:V}]. 

Above, we have pointed out that particle distinguishability is rooted in entanglement between external and internal degrees of freedom. The purity $\tr{\rho_\mathrm{E}^2}$ of the reduced external state quantifies the degree of entanglement and is the basis for our second measure of the wave character, the \emph{normalized purity} 
\begin{align}\label{eq:Pmeasure}
\mathcal{W}_\mathrm{P}=\sqrt{\frac{R}{R-1}\left( \tr{\rho_\mathrm{E}^2}-\frac{1}{R} \right)},
\end{align}
which satisfies $0\leq\mathcal{W}_\mathrm{P}\leq 1$, since the purity is bounded by $1/R\leq \tr{\rho_\mathrm{E}^2} \leq 1$ [compare to Eq.~\eqref{eq:SPP}]. Just as for the previous measure, $\mathcal{W}_\mathrm{P}=0$ corresponds to distinguishable and $\mathcal{W}_\mathrm{P}=1$ to indistinguishable particles.  Note that $\mathcal{W}_\mathrm{P}$ was identified  \cite{Yao-FN-2016} as a coherence measure that is related to the Frobenius norm (or Hilbert-Schmidt norm) $D_\mathrm{HS}(\rho_\mathrm{E},\rho_\mathrm{E}^\mathrm{D})=\sqrt{\tr{|\rho_\mathrm{E}-\rho_\mathrm{E}^\mathrm{D}|^2}}$ of $\rho_\mathrm{E}$ and $\rho_\mathrm{E}^\mathrm{D}$ [see Eqs.~\eqref{eq:rhoE} and~\eqref{eq:rhoEdist}] by $\mathcal{W}_\mathrm{P}=\sqrt{R/(R-1)} D_\mathrm{HS}(\rho_\mathrm{E},\rho_\mathrm{E}^\mathrm{D})$. Moreover, a similar measure for many-particle indistinguishability was proposed in Ref.~\cite{Shchesnovich-PI-2015} [Eq.~(52) there].

As we show in Appendix~\ref{appsec:hierarchy}, like $\mathcal{W}_\mathrm{C}$, the normalized purity $\mathcal{W}_\mathrm{P}$ can be expressed in terms of a sum over the off-diagonal elements of $\rho_\mathrm{E}$. In particular, $\mathcal{W}_\mathrm{C}$ (resp. $\mathcal{W}_\mathrm{P}$) is related to the $L_1$-norm (resp. $L_2$-norm) of a vector whose elements are the off-diagonal elements of $\rho_\mathrm{E}$. In this regard, both measures quantify the ability for the state to display many-particle interference [see Eq.~\eqref{eq:pj-decomp} below]. We further prove in Appendix~\ref{appsec:hierarchy} that these measures obey the following inequality:
\begin{align}\label{eq:hierarchy}
\mathcal{W}_\mathrm{C} \leq \mathcal{W}_\mathrm{P}.
\end{align}
It is worth noting that this inequality does not necessarily saturate for pure internal states. It saturates if and only if all off-diagonal elements~\eqref{eq:rhoEoff-diag} of $\rho_\mathrm{E}$ have equal modulus [see Appendix~\ref{appsec:hierarchy} for details].

\subsubsection{Quantum state discrimination and particle character}\label{sec:particlecharacter}
In section~\ref{sec:reducedstates}, we associated the distinctiveness of the internal states $\rho_\mathrm{I}^{\mu}$ in Eq.~\eqref{eq:rhoImu} with the many-body state's particle character. To quantify this property, we make use of the concept of  \emph{quantum state discrimination}, which we briefly introduce in the following.

Given a quantum state drawn from the set $\{\rho_1,\dots,\rho_k\}$ with corresponding a priori probabilities $\eta_1,\dots,\eta_k$, quantum state discrimination aims at quantifying the ability to discriminate between $\rho_1,\dots,\rho_k$ via a general measurement. Therefore, one considers positive-operator valued measures (POVMs) $\mathcal{M}=\{M_j\}_{j=1}^k$, consisting of positive semidefinite Hermitian operators, which satisfy $\sum_{j=1}^k M_j=\unit$, such that outcome $j$ identifies state $\rho_j$. 
In \emph{minimum error} or \emph{ambiguous} quantum state discrimination (AQSD) \cite{Helstrom-QD-1976,Sacchi-OD-2005,Qiu-MED-2008,Qiu-ME-2010,Spehner-QC-2014}, the outcome of the measurement does not necessarily identify the correct state and one chooses $\mathcal{M}$ such as to maximize the probability of a correct result, leading to the success probability
\begin{align}\label{eq:PA}
P_\mathrm{AQSD}=\max_{\mathcal{M}} \sum_{j=1}^k \eta_j \tr{M_j \rho_j}.
\end{align}
In \emph{unambiguous} quantum state discrimination (UQSD) \cite{Ivanovic-HD-1987,Dieks-OD-1988,Peres-HD-1988,Rudolph-UD-2003,Feng-UD-2004,Spehner-QC-2014}, 
one demands that output $j$ identifies state $\rho_j$ with certainty, which is only possible if one  supplements the POVM $\mathcal{M}$ with a Hermitian operator $M_0$ corresponding to an inconclusive answer. The success probability then reads
\begin{align}\label{eq:PU}
P_\mathrm{UQSD}=\max_{\mathcal{M}} \left( 1-\sum_{j=1}^k \eta_j \tr{M_0\rho_j} \right).
\end{align}
In general, for both success probabilities $P_\mathrm{AQSD}$ and $P_\mathrm{UQSD}$, no exact expressions in terms of distances between $\rho_1,\dots,\rho_k$ are known. However, various upper bounds were derived \cite{Qiu-ME-2010,Spehner-QC-2014}, some of which we utilize in the following.

We now turn towards the quantification of a given many-body state's particle character, with the help of quantum state discrimination. Our aim is to discriminate the internal states $\{\rho_\mathrm{I}^{\mu}\}_{\mu\in\Sigma}$ from Eq.~\eqref{eq:rhoImu}, with equal a priori probabilities $1/R$, by virtue of Eq.~\eqref{eq:rhoI}. For the moment, we consider the discrimination of these internal states as a formal problem, which we make more concrete in Sec.~\ref{sec:permuted} below, where we show its equivalence to the discrimination of common states (including the particles' internal and external degrees of freedom) differing by permutations of the particles. However, for now, let us concentrate on the discrimination of the internal states $\rho_\mathrm{I}^{\mu}$, and use the upper bound on the success probability~\eqref{eq:PA} for AQSD as derived in Ref.~\cite{Qiu-MED-2008}. Given our above definitions, this reads
\begin{align*}
P_\mathrm{AQSD}\leq \frac{1}{2} (1+\mathcal{P}_\mathrm{T}),
\end{align*}
with the trace-distance-based measure
\begin{align}\label{eq:DTT}
\mathcal{P}_\mathrm{T}=\frac{1}{R(R-1)} \sum_{\substack{\mu,\nu\in\Sigma\\ \mu\neq\nu}} D(\rho_\mathrm{I}^{\mu},\rho_\mathrm{I}^{\nu}).
\end{align}
This measure is normalized, $0\leq \mathcal{P}_\mathrm{T} \leq 1$, with the lower bound being reached if all internal states $\rho_\mathrm{I}^{\mu}$ are equal. The upper bound saturates if all pairs of distinct internal states $\rho_\mathrm{I}^{\mu}$ and $\rho_\mathrm{I}^{\nu}$ have orthogonal support. In this regard, the distance measure~\eqref{eq:DTT} quantifies the ability to discriminate particle labelings and, thus, serves as a measure for the particle character. 

A second quantifier of the many-body state's particle character can be motivated by the upper bound on the success probability for UQSD [see Eq.~\eqref{eq:PU}], derived in  \cite{Feng-UD-2004}. For the discrimination of states $\{\rho_\mathrm{I}^{\mu}\}_{\mu\in\Sigma}$ with equal a priori probabilities $1/R$, we have
\begin{align}\label{eq:PUF}
P_\mathrm{UQSD} \leq 1-\mathcal{F},
\end{align}
where the pairwise fidelity measure $\mathcal{F}$ is given by
\begin{align}\label{eq:Ffidel}
\mathcal{F}=\sqrt{  \frac{1}{R(R-1)} \sum_{\substack{\mu,\nu\in\Sigma\\ \mu\neq\nu}} F^2(\rho_\mathrm{I}^{\mu},\rho_\mathrm{I}^{\nu})   }.
\end{align}
From Eq.~\eqref{eq:PUF} it follows that $P_\mathrm{UQSD}\leq\sqrt{1-\mathcal{F}^2}$, which motivates the definition of
\begin{align}\label{eq:DFF}
\mathcal{P}_\mathrm{F}=\sqrt{1-\mathcal{F}^2}
\end{align}
as a measure for the particle character. Similarly to the case of $\mathcal{P}_\mathrm{T}$ from Eq.~\eqref{eq:DTT}, we have $0\leq\mathcal{P}_\mathrm{F}\leq 1$ with $\mathcal{P}_\mathrm{F}=0$ if all internal states $\rho_\mathrm{I}^\mu$ are equal, and $\mathcal{P}_\mathrm{F}=1$ if all pairs of distinct states $\rho_\mathrm{I}^\mu$ have orthogonal support.

With the help of the Fuchs-van de Graaf inequality~\eqref{eq:FvdG}, we obtain a relation reminiscent of Eq.~\eqref{eq:DDinequDouble},
\begin{align}\label{eq:DDinequality}
\mathcal{P}_\mathrm{T}\leq \mathcal{P}_\mathrm{F},
\end{align}
which is proven in Appendix~\ref{appsec:DDinequality}.
Indeed, for $R=2$ different particle labelings (analogous to two mutually exclusive paths $\mathrm{A}$ and $\mathrm{B}$ in the single particle interference scenario discussed in Sec.~\ref{sec:double-slit} above), $\mathcal{P}_\mathrm{T}$ and $\mathcal{P}_\mathrm{F}$ coincide with $\mathcal{P}_\mathrm{t}$ and $\mathcal{P}_\mathrm{f}$ from Eqs.~\eqref{eq:DT} and~\eqref{eq:DF}. However, while $\mathcal{P}_\mathrm{t}=\mathcal{P}_\mathrm{f}$ for pure states, in general Eq.~\eqref{eq:DDinequality} does not saturate for pure internal states in the case $R>2$.

\subsection{Wave-particle duality}
\label{sec:duality}
So far, for a state of many partially distinguishable particles, we related the measures $\mathcal{W}_\mathrm{C}$ and $\mathcal{W}_\mathrm{P}$ [see Eqs.~\eqref{eq:Cmeasure} and~\eqref{eq:Pmeasure}] to its many-particle wave character, and $\mathcal{P}_\mathrm{T}$ and $\mathcal{P}_\mathrm{F}$ [see Eqs.~\eqref{eq:DTT} and~\eqref{eq:DFF}] to its particle character. These measures quantify the ability of particles to display many-particle interference on the one hand, and the possibility of individually identifying and tracking them, on the other hand. Wave-particle duality relations state that both properties cannot be fully realized in the same state: the amount of wave character limits the amount of particle character, and vice versa. For $\mathcal{P}_\mathrm{F}$ and $\mathcal{W}_\mathrm{P}$, this is quantitatively expressed by
\begin{align}\label{eq:CompMany}
\mathcal{P}_\mathrm{F}^2+\mathcal{W}_\mathrm{P}^2 \leq 1,
\end{align}
which we prove in Appendix~\ref{appsec:CompProof}. This constitutes a wave-particle duality relation since it saturates for pure internal states. By the hierarchies~\eqref{eq:hierarchy} and~\eqref{eq:DDinequality} (which do not necessarily saturate for pure internal states), we additionally find complementarity relations between all combinations of the above defined wave and particle measures:
\begin{align}\label{eq:CompManyAll}
\mathcal{P}^2+\mathcal{W}^2 \leq 1,
\end{align}
for $\mathcal{P}\in\{\mathcal{P}_\mathrm{T},\mathcal{P}_\mathrm{F}\}$ and $\mathcal{W}\in\{\mathcal{W}_\mathrm{C},\mathcal{W}_\mathrm{P}\}$. Interestingly, the complementarity relation~\eqref{eq:CompManyAll} saturates for pure internal states for the wave measure $\mathcal{W}_\mathrm{P}$, but not for $\mathcal{W}_\mathrm{C}$. While the former quantifies the correlations between internal and external degrees of freedom independently of the chosen basis via their entanglement, the latter measures these correlations via the coherence of $\rho_\mathrm{E}$ in the chosen external basis. Therefore, the entanglement-based measure $\mathcal{W}_\mathrm{P}$ seems to have a more fundamental status, while the coherence-based one $\mathcal{W}_\mathrm{C}$ might be more suited to describe measurements performed in a specific basis.

\begin{figure}[t]
\centering
\includegraphics[width=\linewidth]{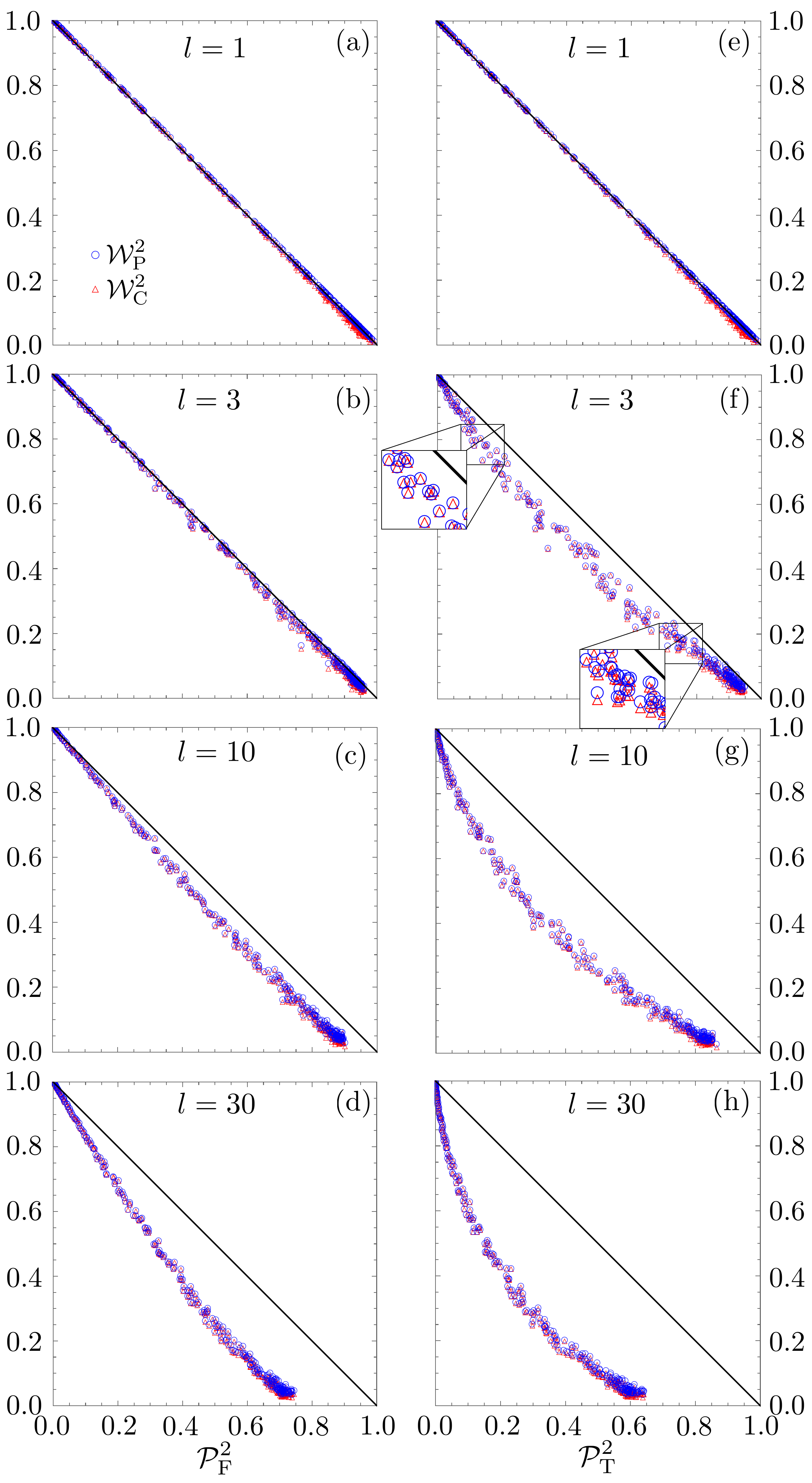}
\caption{Complementarity relations for $300$ randomly generated states of three partially distinguishable particles, with $l$ referring to the number of mixed internal states [see main text for details]. Panels (a-d) show the wave character measures $\mathcal{W}_\mathrm{P}^2$ (blue circles) and $\mathcal{W}_\mathrm{C}^2$ (red triangles) plotted against the particle character quantifier $\mathcal{P}_\mathrm{F}^2$, while in panels (e-h) they are plotted against $\mathcal{P}_\mathrm{T}^2$. In all panels, the solid black line corresponds to the upper bound according to~\eqref{eq:CompManyAll}.
\label{fig:simulations}}
\end{figure}

All complementarity relations~\eqref{eq:CompManyAll} and their dependence on the mixedness of the internal state are illustrated via a numerical example in Fig.~\ref{fig:simulations}. We consider $N=3$ particles occupying distinct modes, with external and internal single-particle Hilbert spaces of dimension $n=m=4$. We generate $300$ random states of partially distinguishable particles by mixing $l=1,3,10$ or $30$ different pure internal states~\eqref{eq:Omegamusym} [see Eq.~\eqref{eq:rho0}]. The probabilities $q_j$ in~\eqref{eq:rho0} are randomly chosen according to a uniform probability distribution in the range $q_j\in[0,1]$, and normalized such that $\sum_{j=1}^l q_j=1$. Many-particle distinguishability is encoded in the coefficients $C_{\vec{\mathcal{I}}}^{(j)}=r_{\vec{\mathcal{I}}}^{(j)}\exp(i\varphi_{\vec{\mathcal{I}}}^{(j)})$ -- which enter through~\eqref{eq:Omegamusym}. To fairly distribute the $300$ generated states over all possible levels of partial distinguishability, for the internal states of the $k$th state (with $0\leq k \leq 300$), we uniformly pick $r_{\vec{\mathcal{I}}}^{(j)}\in[1-k/300,1]$ and $\varphi_{\vec{\mathcal{I}}}^{(j)}\in[-\pi k/300,\pi k/300]$ for each $\vec{\mathcal{I}}$ and each pure state $j$. This results in $l m^N=l 4^3$ different coefficients $C_{\vec{\mathcal{I}}}^{(j)}$, which are then appropriately normalized, such that $\sum_{\vec{\mathcal{I}}} |C_{\vec{\mathcal{I}}}^{(j)}|^2=1$ for each $j=1,\dots,l$. Note that no symmetrization of the coefficients is required since we consider at most singly occupied modes.

 In Figs.~\ref{fig:simulations}(a-d) and \ref{fig:simulations}(e-h), we plot the squared wave character quantifiers $\mathcal{W}_\mathrm{C}^2$ and $\mathcal{W}_\mathrm{P}^2$ against the squared particle character quantifiers $\mathcal{P}_\mathrm{F}^2$ and  $\mathcal{P}_\mathrm{T}^2$, for $\l=1,3,10,$ and $30$. Pure internal states  ($l=1$) come close to saturating the upper bound for all complementarity relations~\eqref{eq:CompManyAll}, as shown in Figs.~\ref{fig:simulations}(a) and (e). In particular, the saturation of Eq.~\eqref{eq:CompMany} is evident from Fig.~\ref{fig:simulations}(a). However, for stronger mixing of the internal states, i.e. for increasing $l$, the sum of the squared quantifiers of complementary many-particle properties tends to move away from the upper bound which, by Eq.~\eqref{eq:DDinequality}, is stronger for $\mathcal{P}_\mathrm{T}^2$ as compared to  $\mathcal{P}_\mathrm{F}^2$ [compare Figs.~\ref{fig:simulations}(a-d) with Figs.~\ref{fig:simulations}(e-h)]. Note that while full wave character quantifiers imply vanishing particle character quantifiers, the converse is not true. Indeed, mixing of internal states can reduce the particle character quantifiers [see the discussion below Eq.~\eqref{eq:rhoImu}], such that states with a low wave character can also have low particle character. For all sampled states, we find very similar values for both wave character quantifiers. As shown by the zooms in Fig.~\eqref{fig:simulations}(f), the difference between $\mathcal{W}_\mathrm{P}^2$ and $\mathcal{W}_\mathrm{C}^2$ tends to increase with more pronounced particle character, which can be traced back to inhomogeneities in the moduli of the off-diagonal elements of $\rho_\mathrm{E}$ [see Appendix~\ref{appsec:hierarchy}].

Let us now return to the analogy between many-particle complementarity as summarized by~\eqref{eq:CompManyAll} and the double-slit experiment discussed in Sec.~\ref{sec:double-slit}. First of all, the common state~\eqref{eq:rhoPD} of particle and detector is structurally very similar to the many-body state~\eqref{eq:rho} of external and internal degrees of freedom. In both cases, the weaker the entanglement between the subsystems, the more pronounced the wave character of the state. This observation has its direct counterpart in the congruent structure of the wave character quantifiers in Eqs.~\eqref{eq:SPP} and~\eqref{eq:Pmeasure}. Consistently, the wave character is related to the coherences of the reduced single-particle state by~\eqref{eq:V}, and of the reduced many-particle external state by~\eqref{eq:Cmeasure}. The latter, however, leaves room for a much more subtle quantum-classical transition than the former, due to the many degrees of freedom involved. 

Likewise, the particle character is determined by the ability to discriminate the states~\eqref{eq:rhoD} of the which-path detectors in the single-particle case and the many-body internal states~\eqref{eq:rhoI} in case of partially distinguishable particles. To quantify the particle character, concepts imported from quantum state discrimination lead to a generalization of Eqs.~\eqref{eq:DT} and~\eqref{eq:DF} by ~\eqref{eq:DTT} and~\eqref{eq:DFF}. This deep analogy then results in the many-body generalization~\eqref{eq:CompManyAll} of the single particle duality relation~\eqref{eq:Englert2}. For the sake of clarity, all character measures and complementarity relations presented in Secs.~\ref{sec:double-slit} and~\ref{sec:pd-particles} are summarized in Table~\ref{tab:SpMpComp}.

\begin{table}[t]
\begin{ruledtabular}
\begin{tabular}{c c c c}
Particle  & Wave  & Complementarity  & Saturation  \\ 
measures & measures & relations & for pure states\\ \hline 
\multicolumn{4}{c}{\multirow{2}{*}{Single-particle double-slit}}\\  \\ 
\rule{0pt}{10pt} \multirow{2}{*}{$\mathcal{P}_\mathrm{t}\leq\mathcal{P}_\mathrm{f}$} &  \multirow{2}{*}{$\mathcal{V}$    } & $\mathcal{P}_\mathrm{f}^2+\mathcal{V}^2 \leq 1$ & yes\\ \cline{3-4} 
\rule{0pt}{10pt}& & $\mathcal{P}_\mathrm{t}^2+\mathcal{V}^2 \leq 1$ & yes\\
\hline
\multicolumn{4}{c}{\multirow{2}{*}{Many-particle states}}\\ \\ 
\rule{0pt}{10pt} \multirow{4}{*}{$\mathcal{P}_\mathrm{T}\leq \mathcal{P}_\mathrm{F}$} &  \multirow{4}{*}{$\mathcal{W}_\mathrm{C}\leq \mathcal{W}_\mathrm{P}$    } & $\mathcal{P}_\mathrm{F}^2+\mathcal{W}_\mathrm{P}^2 \leq 1$ & yes\\ \cline{3-4} 
\rule{0pt}{10pt} & & $\mathcal{P}_\mathrm{F}^2+\mathcal{W}_\mathrm{C}^2 \leq 1$ & no\\ \cline{3-4} 
\rule{0pt}{10pt} & & $\mathcal{P}_\mathrm{T}^2+\mathcal{W}_\mathrm{P}^2 \leq 1$ & no\\ \cline{3-4} 
\rule{0pt}{10pt} & & $\mathcal{P}_\mathrm{T}^2+\mathcal{W}_\mathrm{C}^2 \leq 1$ & no\\ 
\end{tabular}
\caption{Overview of complementarity relations for the single particle double-slit experiment, and for states of many partially distinguishable particles.}
\label{tab:SpMpComp}
\end{ruledtabular}
\end{table}

\section{Wave-particle duality in the Interference of Many Particles}
\label{sec:visibilities}
In the previous section, we considered states of partially distinguishable particles and associated the coherence and the purity of their reduced external states $\rho_\mathrm{E}$ with their wave character, arguing that both properties give rise to many-particle interference. We now substantiate this claim by investigating the outcome of interference experiments performed with partially distinguishable particles.

\begin{figure}[t]
\centering
\includegraphics[width=\linewidth]{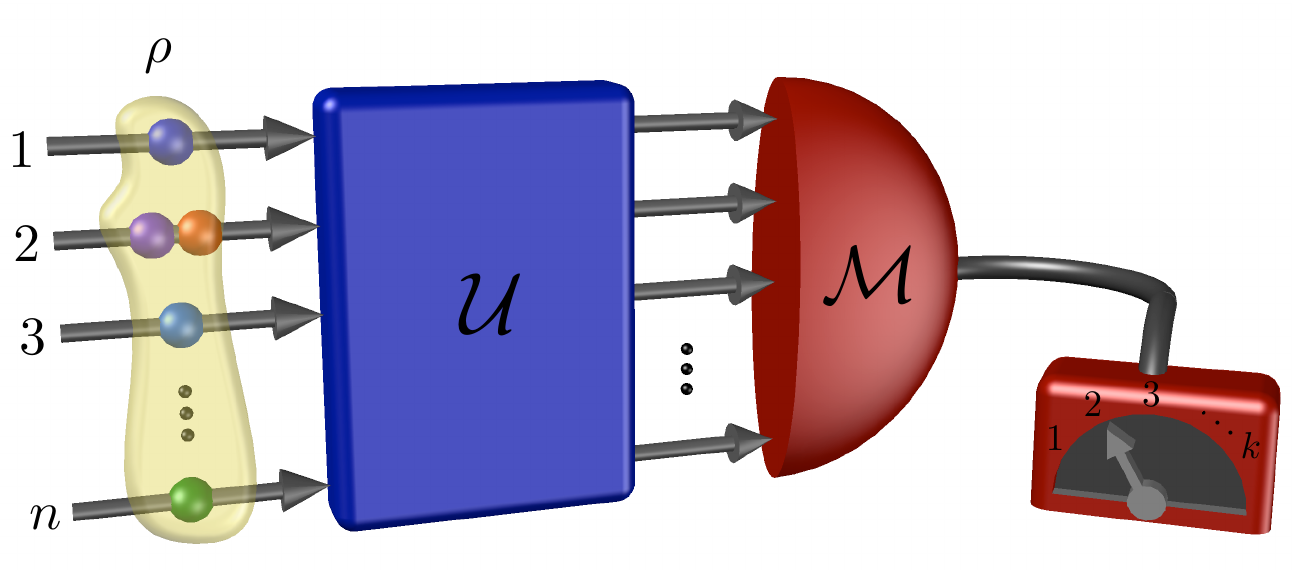}
\caption{\label{fig:MPExperiment}General setting of a many-particle interference experiment with a state $\rho$ of $N$ partially distinguishable particles in $n$ modes (colored balls covered by a yellow envelope illustrating correlations in the internal degrees of freedom). The external state $\rho_\mathrm{E}$ evolves according to a many-particle, possibly interacting, unitary $\mathcal{U}\in\mathrm{U}(n^N)$ (illustrated in blue) and is measured by a POVM $\mathcal{M}$ with outcomes $1, \dots, k$ (illustrated in red). As a consequence of wave-particle duality, the visibility of many-particle interference is bounded by the particles' degree of distinguishability.}

\end{figure}

The experimental arrangement under consideration is depicted in Fig.~\ref{fig:MPExperiment} and consists of a generic state~\eqref{eq:rho} of $N$ partially distinguishable particles distributed over $n$ modes, which undergoes coherent evolution of its external degrees of freedom while the internal state remains unaffected. We thus consider the evolution of the reduced external state $\rho_\mathrm{E}$ under an arbitrary many-particle unitary $\mathcal{U}$, chosen from the unitary group $\mathrm{U}(n^N)$. Note that, since $\mathcal{U}$ acts on many identical particles, it must commute with all particle permutation operators. Further note that, in general, $\mathcal{U}$ describes an interacting evolution, and not only a unitary mapping of the input modes to the output modes -- thus the degree of $\mathcal{U}$ is exponential in the particle number. After the evolution, the resulting state $\mathcal{U}\rho_\mathrm{E}\mathcal{U}^\dagger$ is measured with the help of a POVM $\mathcal{M}=\{M_j\}_{j=1}^k$. Similar to $\mathcal{U}$, the operators $M_j$ must commute with all particle permutation operators. In principle one could absorb the unitary evolution into the measurement $\mathcal{M}$, however, for the sake of clarity, we consider evolution and detection stages separately. 

Let us assume that the POVM $\mathcal{M}=\{M_j\}_{j=1}^k$ has $k$ distinct outcomes and results in the associated counting statistics $P_\mathcal{M}=\{p(j)\}_{j=1}^k$ for partially distinguishable particles. Here, $p(j)=\tr{M_j \mathcal{U}\rho_\mathrm{E} \mathcal{U}^\dagger}$ is the probability of outcome $j\in\{1,\dots,k\}$, which can be decomposed as
\begin{align}\label{eq:pj-decomp}
p(j)=p^\mathrm{D}(j)+ \sum_{\substack{\mu,\nu \in \Sigma \\ \mu\neq \nu}} [\rho_\mathrm{E}]_{\mu,\nu} \bra{\vec{E}_\nu} \mathcal{U}^\dagger M_j \mathcal{U} \ket{\vec{E}_\mu},
\end{align}
with $p^\mathrm{D}(j)=\tr{M_j \mathcal{U}\rho_\mathrm{E}^\mathrm{D} \mathcal{U}^\dagger}$ the probability in the case of fully distinguishable particles, and the second term accounting for many-particle interference as governed by the coherences (off-diagonal elements) $[\rho_\mathrm{E}]_{\mu,\nu}$ from Eq.~\eqref{eq:rhoEoff-diag}. Note that, by Eq.~\eqref{eq:pj-decomp}, the wave character measures $\mathcal{W}_\mathrm{C}$ and $\mathcal{W}_\mathrm{P}$ from Eqs.~\eqref{eq:Cmeasure} and~\eqref{eq:Pmeasure} [see also Eq.~\eqref{eq:PmeasureApp}] ultimately dictate the many-particle state's ability to interfere.

In the following, we consider distances between such counting statistics on output. We make use of the classical analogues of trace distance $D(\rho,\sigma)$ and quantum fidelity $F(\rho,\sigma)$, known respectively as the \emph{Kolmogorov distance} (or \emph{$L_1$ distance}), 
\begin{align}\label{eq:Dclas}
D(P^\mathrm{A}_\mathcal{M},P^\mathrm{B}_\mathcal{M})=\frac{1}{2}\sum_{j=1}^k\abs{p_\mathrm{A}(j)-p_\mathrm{B}(j)},
\end{align}
and \emph{Bhattacharyya coefficient} (or \emph{fidelity}),
\begin{align}\label{eq:BhattaCoef}
F(P^\mathrm{A}_\mathcal{M},P^\mathrm{B}_\mathcal{M})=\sum_{j=1}^k\sqrt{p_\mathrm{A}(j)p_\mathrm{B}(j)},
\end{align}
which both take values between $0$ and $1$ \cite{Fuchs-CD-1999,Nielsen-QC-2011}. 
These measures are related to the trace distance and quantum fidelity by an optimization over all POVMs:
if $\rho_\mathrm{A}$ and $\rho_\mathrm{B}$ are two states leading to distributions $P^\mathrm{A}_\mathcal{M}$ and $P_\mathcal{M}^\mathrm{B}$, then \cite{Fuchs-CD-1999}
\begin{align}\label{eq:relationD}
D(\rho_\mathrm{A},\rho_\mathrm{B})=\max_{\mathcal{M}} D(P^\mathrm{A}_\mathcal{M},P_\mathcal{M}^\mathrm{B})
\end{align}
and
\begin{align}\label{eq:relationF}
F(\rho_\mathrm{A},\rho_\mathrm{B})=\min_{\mathcal{M}} F(P^\mathrm{A}_\mathcal{M},P_\mathcal{M}^\mathrm{B}).
\end{align}
In analogy to their quantum counterparts [cf. Eq.~\eqref{eq:FvdG}], these measures obey the Fuchs-van de Graaf inequality  \cite{Fuchs-CD-1999} 
\begin{align}\label{eq:DistanceIneqClas}
D(P^\mathrm{A}_\mathcal{M},P_\mathcal{M}^\mathrm{B})\leq \sqrt{1-F^2(P^\mathrm{A}_\mathcal{M},P_\mathcal{M}^\mathrm{B})}.
\end{align}

\subsection{Permutations of the internal states}
\label{sec:permuted}
In Sec.~\ref{sec:particlecharacter}, we defined the measures $\mathcal{P}_\mathrm{T}$ and $\mathcal{P}_\mathrm{F}$ [see Eqs.~\eqref{eq:DTT} and~\eqref{eq:DFF}] to quantify the many-body state's particle character. These measures are based on the formal discrimination of different particle labelings $\mu\in\Sigma$ by comparison of the associated internal states $\rho_\mathrm{I}^{\mu}$ from Eq.~\eqref{eq:rhoImu}. 
We now show that different particle labelings can equivalently be discriminated by comparing \emph{common states}, of external \emph{and} internal degrees of freedom, that differ by an initial permutation of the internal states. Therefore, instead of the unpermuted internal many-particle states $\ket{\Omega^{(j)}}$ from Eq.~\eqref{eq:Omega}, we consider the internal states $\ket{\Omega_\kappa^{(j)}}$ permuted according to $\kappa\in\Sigma$ [see Eq.~\eqref{eq:Omegamusym}]. While the unpermuted common state $\rho$ of external and internal degrees of freedom is given in Eq.~\eqref{eq:rho}, the \emph{permuted states} read
\begin{align}\label{eq:rhomu}
\rho^\kappa=\frac{1}{R} \sum_{\mu,\nu\in\Sigma} (-1)^{\mu\nu}_\mathrm{B(F)} \ket{\vec{E}_\mu}\bra{\vec{E}_\nu} \otimes \sum_j q_j \ket{\Omega_{\kappa\mu}^{(j)}}\bra{\Omega_{\kappa\nu}^{(j)}}.
\end{align}
Here, the subscript $\kappa\mu$ in $\ket{\Omega_{\kappa\mu}^{(j)}}$ refers to a composition of permutations, with permutation $\kappa$ and $\mu$ arising due to the initially permuted internal states and the symmetrization of the many-particle state, respectively.

In Appendix~\ref{appsec:Permuted}, we prove the equality of the distances 
\begin{align}\label{eq:Dequiv}
D(\rho^{\kappa},\rho^{\tau})=D(\rho_\mathrm{I}^{\kappa},\rho_\mathrm{I}^{\tau}),
\end{align}
and of the fidelities
\begin{align}\label{eq:Fequiv}
F(\rho^{\kappa},\rho^{\tau})=F(\rho_\mathrm{I}^{\kappa},\rho_\mathrm{I}^{\tau}),
\end{align}
with $\rho^{\kappa}$ and $\rho_\mathrm{I}^{\kappa}$ from  Eq.~\eqref{eq:rhomu} and~\eqref{eq:rhoImu}, respectively. Thus, the discrimination of different particle labelings $\kappa$ can be performed equally well by comparison of the internal states $\rho_\mathrm{I}^{\kappa}$, or of the permuted common states $\rho^{\kappa}$. We therefore investigate how the outcomes of interference experiments as sketched in Fig.~\eqref{fig:MPExperiment} differ for permuted states. This will lead us to classical counterparts of the measures $\mathcal{P}_\mathrm{T}$ and $\mathcal{P}_\mathrm{F}$ of the particle character, evaluated on the output counting statistics, which also obey complementarity relations of the form~\eqref{eq:CompManyAll}.

Let us denote by $\rho_\mathrm{E}^{\kappa}$ the reduced external state of the permuted state, $\rho_\mathrm{E}^{\kappa}=\trp{I}{\rho^{\kappa}}$, and by $P_\mathcal{M}^\kappa$ the probability distribution obtained when $\rho^{\kappa}$ is used as input in the experiment depicted in Fig.~\ref{fig:MPExperiment}. As we prove in Appendix~\ref{appsec:Cmunu}, for two permutations $\kappa \neq \tau$, the classical distances between the corresponding output probability distributions are bounded by the magnitude of the corresponding off-diagonal element of $\rho_\mathrm{E}$,
\begin{align}\label{eq:Cmunuclass}
\nonumber D(P_\mathcal{M}^\kappa,P_\mathcal{M}^\tau) \leq & \sqrt{1-F^2(P_\mathcal{M}^\kappa,P_\mathcal{M}^\tau)}\\
\leq & \sqrt{1-R^2 \abs{\bra{\vec{E}_\kappa}\rho_\mathrm{E}\ket{\vec{E}_\tau}}^2}.
\end{align}
Thus, for external states $\rho_\mathrm{E}$ with large coherences, $|\bra{\vec{E}_\kappa}\rho_\mathrm{E}\ket{\vec{E}_\tau}| \approx 1/R$, permuted input states lead to similar output probability distributions.

Given this observation, it is natural to consider all pairwise differences in the output probability distributions $P_\mathcal{M}^\kappa,\ P_\mathcal{M}^\tau, \ \kappa\neq\tau$. We therefore define classical analogues of $\mathcal{P}_\mathrm{T}$ and $\mathcal{P}_\mathrm{F}$ from Eqs.~\eqref{eq:DTT} and~\eqref{eq:DFF}:
\begin{align}\label{eq:DcalT}
\mathscr{P}_\mathrm{T}=\frac{1}{R(R-1)} \sum_{\substack{\kappa,\tau\in\Sigma\\ \kappa\neq\tau}} D(P_\mathcal{M}^\kappa,P_\mathcal{M}^\tau),
\end{align}
and
\begin{align}\label{eq:DcalF}
\mathscr{P}_\mathrm{F}=\sqrt{1-  \frac{1}{R(R-1)} \sum_{\substack{\kappa,\tau\in\Sigma\\ \kappa\neq\tau}} F^2(P_\mathcal{M}^\kappa,P_\mathcal{M}^\tau)   },
\end{align}
with $0\leq \mathscr{P}_\mathrm{T},\mathscr{P}_\mathrm{F} \leq 1$. Similarly to Eqs.~\eqref{eq:DDinequDouble} and~\eqref{eq:DDinequality}, these measures obey
\begin{align}\label{eq:DcalDcal}
\mathscr{P}_\mathrm{T} \leq \mathscr{P}_\mathrm{F},
\end{align}
and they are bounded from above by their quantum counterparts,
\begin{align}\label{eq:DTDFinequ}
\mathscr{P}_\mathrm{T} \leq \mathcal{P}_\mathrm{T}  \quad \text{and} \quad \mathscr{P}_\mathrm{F} \leq \mathcal{P}_\mathrm{F}.
\end{align}
Equation~\eqref{eq:DcalDcal} and~\eqref{eq:DTDFinequ} are proven in Appendices~\ref{appsec:DcalDcal} and~\ref{appsec:CompDcal}, respectively. In consideration of Eq.~\eqref{eq:CompManyAll}, the inequalities in~\eqref{eq:DTDFinequ} directly lead to the complementarity relations
\begin{align}\label{eq:CompDcal} 
\mathscr{P}^2+\mathcal{W}^2 \leq 1,
\end{align}
for  $\mathscr{P}\in\{\mathscr{P}_\mathrm{T},\mathscr{P}_\mathrm{F}\}$ and $\mathcal{W}\in\{\mathcal{W}_\mathrm{C},\mathcal{W}_\mathrm{P}\}$. In contrast to Eq.~\eqref{eq:CompManyAll}, these relations use experimental outcomes in the external degrees of freedom to quantify the particle character. In the case of indistinguishable particles, the state remains invariant under permutations of the particle's internal degrees of freedom and thus $P_\mathcal{M}^\kappa$ is the same for all permutations $\kappa$, such that $\mathscr{P}_\mathrm{T} = \mathscr{P}_\mathrm{F} =0$ [see Eqs.~\eqref{eq:DcalT} and~\eqref{eq:DcalF}]. On the other hand, permuting partially distinguishable particles in the input can change the output counting statistics, which then leads to non-vanishing measures $\mathscr{P}_\mathrm{T}$ and $ \mathscr{P}_\mathrm{F}$.

\subsection{Partially distinguishable vs. indistinguishable particles}
\label{sec:pdistVSindist}
When considering experiments with partially distinguishable particles, it is common to compare the output distribution against the extreme distributions obtained with fully distinguishable or fully indistinguishable particles. In this way, one can define generalized visibilities of many-particle interference. Here, we start by comparing the output probability distribution $P_\mathcal{M}$ to the one obtained with strictly indistinguishable bosons (fermions) $P^{\mathrm{B}(\mathrm{F})}_\mathcal{M}$. This question was addressed in Ref.~\cite{Shchesnovich-TB-2015} in the context of Boson Sampling, that is for non-interacting bosons and  (possibly imperfect) particle number measurements. Here we generalize the result of \cite{Shchesnovich-TB-2015} to both, bosons and fermions, interacting evolutions, and arbitrary measurements.

We start by noting that $\ket{\psi_{\mathrm{B}(\mathrm{F})}}$ from Eq.~\eqref{eq:rhoEBFpure} is an eigenvector of $\rho_\mathrm{E}$ from Eq.~\eqref{eq:rhoE}, with eigenvalue $\lambda_{\mathrm{B(F)}}= F^2(\rho_\mathrm{E}^{\mathrm{B(F)}},\rho_\mathrm{E})$:
\begin{align}\label{eq:eigenequation}
\rho_\mathrm{E}\ket{\psi_{\mathrm{B}(\mathrm{F})}}=  \lambda_{\mathrm{B(F)}} \ket{\psi_{\mathrm{B}(\mathrm{F})}}.
\end{align}
This is proven in Appendix~\ref{appsec:Eigenequation}. As we further show in Appendix~\ref{appsec:FvGspecial}, Eq.~\eqref{eq:eigenequation} leads to
\begin{align}\label{eq:FvGspecial}
D(\rho_\mathrm{E}^{\mathrm{B}(\mathrm{F})},\rho_\mathrm{E})+F^2(\rho_\mathrm{E}^{\mathrm{B}(\mathrm{F})},\rho_\mathrm{E})=1.
\end{align}
Given the relation~\eqref{eq:relationD} between the trace distance and the Kolmogorov distance, as well as the invariance of the trace distance under unitary transformations \cite{Nielsen-QC-2011}, we find
\begin{align}\label{eq:Vis01}
D(P^{\mathrm{B}(\mathrm{F})}_\mathcal{M},P_\mathcal{M})+F^2(\rho_\mathrm{E}^{\mathrm{B}(\mathrm{F})},\rho_\mathrm{E}) \leq 1,
\end{align}
with the bound saturating for an optimal measurement [see Appendix~\ref{appsec:Vis01} for details].
The difference between the outcomes of experiments performed with partially distinguishable and with indistinguishable particles is thus rather intuitively constrained by the similarity of the input state to the state of ideal bosons or fermions, as measured by the fidelity $F^2(\rho_\mathrm{E}^{\mathrm{B}(\mathrm{F})},\rho_\mathrm{E})$. Furthermore, this fidelity is related to the coherences of $\rho_\mathrm{E}$, and therefore to the state's wave character, through
\begin{align}
F^2(\rho_\mathrm{E}^{\mathrm{B}(\mathrm{F})},\rho_\mathrm{E})=\frac{1}{R} + \frac{1}{R}\sum_{\substack{\mu,\nu\in\Sigma\\ \mu\neq\nu}} (-1)^{\mu\nu}_\mathrm{B(F)} \bra{\vec{E}_\mu} \rho_\mathrm{E} \ket{\vec{E}_\nu},
\end{align}
as can be seen by singling out the terms with $\mu=\nu$ in Eq.~\eqref{eq:fidelitysq} and recognizing the coherences from Eq.~\eqref{eq:rhoE}. Let us stress that while both measures that enter Eq.~\eqref{eq:Vis01} generally vary between zero and unity, in the case of fully distinguishable particles, with the external state $\rho_\mathrm{E}^\mathrm{D}$, we have $F^2(\rho_\mathrm{E}^{\mathrm{B}(\mathrm{F})},\rho_\mathrm{E}^{\mathrm{D}})=1/R$ and, in turn, $D(P^{\mathrm{B}(\mathrm{F})}_\mathcal{M},P^\mathrm{D}_\mathcal{M})\leq 1-1/R$. The Kolmogorov distance therefore does not reach its maximum value for distinguishable particles.

Under the assumptions of non-interacting particles and perfect particle number measurement, the unitary evolution is constrained to the form $\mathcal{U}=u^{\otimes N}$ with $u\in \mathrm{U}(n)$, and the particle number measurement is performed by the projectors $M_{\vec{S}}=\sum_{\mu\in\Sigma(\vec{S})}\ket{\vec{E}_\mu(\vec{S})}\bra{\vec{E}_\mu(\vec{S})}$, where $\vec{S}$ is the output mode occupation list, defined analogously to $\vec{R}$ in Sec.~\ref{sec:pd-particles-intro}. This particular scenario has been addressed in Ref.~\cite{Shchesnovich-TB-2015} and is covered by  Eq.~\eqref{eq:Vis01}. In this regard, we can identify $p_s$ in Eq.~(12) of~\cite{Shchesnovich-TB-2015} with $F^2(\rho_\mathrm{E}^{\mathrm{B}},\rho_\mathrm{E})$. We also note that similar considerations were made under a group-theoretical perspective in Ref.~\cite{Stanisic-DD-2018} [e.g. compare our Eq.~\eqref{eq:FvGspecial} to Eq.~(66) in~\cite{Stanisic-DD-2018}].

\subsection{Partially distinguishable vs. fully distinguishable particles}
\label{sec:pdistVSdist}
We now compare the outcomes of experiments carried out with partially distinguishable particles to those obtained with fully distinguishable particles. To this end, we utilize the Kolmogorov distance~\eqref{eq:Dclas} and the Bhattacharyya coefficient~\eqref{eq:BhattaCoef}, and define the visibilities
\begin{align}\label{eq:VcT}
\mathscr{V}_\mathrm{T}=\frac{R}{R-1} D(P_\mathcal{M}^\mathrm{D},P_\mathcal{M})
\end{align}
and
\begin{align}\label{eq:Vc}
\mathscr{V}_\mathrm{F}=\frac{R}{R-1}\bigg(1-F^2(P^\mathrm{D}_\mathcal{M},P_\mathcal{M})\bigg).
\end{align} 
These visibility measures quantify the interference contrast, they are normalized, $0\leq \mathscr{V}_\mathrm{T},\mathscr{V}_\mathrm{F} \leq 1$, and yield $\mathscr{V}_\mathrm{T}=\mathscr{V}_\mathrm{F}=0$ for distinguishable particles. For indistinguishable particles one finds $\mathscr{V}_\mathrm{T},\mathscr{V}_\mathrm{F}\leq 1$, with the saturation $\mathscr{V}_\mathrm{T}= 1$  (resp. $\mathscr{V}_\mathrm{F}= 1$) in the case of an optimal measurement. 

As a consequence of the relations between quantum and classical trace distance and fidelity [cf. Eqs.~\eqref{eq:relationD} and~\eqref{eq:relationF}], $\mathscr{V}_\mathrm{T}$ and $\mathscr{V}_\mathrm{F}$ are smaller than their quantum counterpart, which we use to define the distinguishability measures $\mathcal{D}_\mathrm{T}$ and $\mathcal{D}_\mathrm{F}$,
\begin{align}\label{eq:VisT}
\mathscr{V}_\mathrm{T}\leq\frac{R}{R-1}D(\rho_\mathrm{E}^{\mathrm{D}},\rho_\mathrm{E} )=1-\mathcal{D}_\mathrm{T},
\end{align}
and
\begin{align}\label{eq:Dc}
\mathscr{V}_\mathrm{F}\leq\frac{R}{R-1}\bigg(1-F^2(\rho_\mathrm{E}^{\mathrm{D}},\rho_\mathrm{E} )\bigg)= 1-\mathcal{D}_\mathrm{F},
\end{align} 
respectively, with $\rho_\mathrm{E}^{\mathrm{D}}$ from Eq.~\eqref{eq:rhoEdist}. The inequalities in~\eqref{eq:VisT} and~\eqref{eq:Dc} saturate for an optimal measurement, such as a projection onto the the eigenstates of $\mathcal{U}\rho_\mathrm{E} \mathcal{U}^\dagger$ [see e.g. Secs.~9.2.1 and 9.2.2 in \cite{Nielsen-QC-2011}].

Given that $1/R\leq F^2(\rho_\mathrm{E}^{\mathrm{D}},\rho_\mathrm{E})\leq 1$ (resp. $0\leq D(\rho_\mathrm{E}^{\mathrm{D}},\rho_\mathrm{E})\leq (R-1)/R $), with the lower (resp. upper) bound reached when $\rho_\mathrm{E}=\rho_\mathrm{E}^{\mathrm{B(F)}}$,  we have $0\leq \mathcal{D}_\mathrm{T},\mathcal{D}_\mathrm{F} \leq 1$, with the lower bound saturating for indistinguishable particles, and the upper bound for distinguishable particles. Therefore, Eqs.~\eqref{eq:VisT} and~\eqref{eq:Dc} entail the complementarity relations
\begin{align}\label{eq:Vis02-T}
\mathcal{D}_\mathrm{T}+\mathscr{V}_\mathrm{T} \leq 1,
\end{align} 
and
\begin{align}\label{eq:Vis02}
\mathcal{D}_\mathrm{F}+\mathscr{V}_\mathrm{F} \leq 1,
\end{align} 
which provide a bound on the visibility based on the distinguishability of the particles, as measured by $\mathcal{D}_\mathrm{T}$ and $\mathcal{D}_\mathrm{F}$, regardless of the exact form of $\rho_\mathrm{E}$, which depends, for example, on the particle type. In other words, a given level of visibility can only be achieved by states that are sufficiently distant from the state of distinguishable particles.

Interestingly, the distinguishability measures $\mathcal{D}_\mathrm{T}$ and $\mathcal{D}_\mathrm{F}$ are also connected  to the previously defined wave character measures~\eqref{eq:Cmeasure} and~\eqref{eq:Pmeasure} by the complementarity relation
\begin{align}\label{eq:Connect}
\mathcal{D}^2+\mathcal{W}^2 \leq 1,
\end{align} 
for $\mathcal{D}\in\{ \mathcal{D}_\mathrm{T},\mathcal{D}_\mathrm{F} \}$ and $\mathcal{W}\in\{\mathcal{W}_\mathrm{C},\mathcal{W}_\mathrm{P}\}$. Inequality~\eqref{eq:Connect} is proven in Appendix~\ref{appsec:Connect}. Although Eq.~\eqref{eq:Connect} does not explicitly refer to outcomes of experiments, it once more highlights the suppression of the wave character by particle distinguishability. For a better overview, Table~\ref{tab:vis} summarizes the relations developed in the present section. 

\begin{table}[t]
\begin{ruledtabular}
\begin{tabular}{c c c c}
\multicolumn{2}{c}{\multirow{2}{*}{Measures} } &Complemen- & Saturation for opti-  \\ 
 &   &tarity relations & mal measurement\\ \hline
\multicolumn{4}{c}{\multirow{2}{*}{Permuted input particles}}\\ \\  
\rule{0pt}{10pt} \multirow{4}{*}{$\mathscr{P}_\mathrm{T}\leq \mathscr{P}_\mathrm{F}$ } &  \multirow{4}{*}{$\mathcal{W}_\mathrm{C}\leq \mathcal{W}_\mathrm{P}$   } & $\mathscr{P}_\mathrm{F}^2+\mathcal{W}_\mathrm{P}^2 \leq 1$ & no\\ \cline{3-4} 
\rule{0pt}{10pt} & & $\mathscr{P}_\mathrm{F}^2+\mathcal{W}_\mathrm{C}^2 \leq 1$ & no\\ \cline{3-4} 
\rule{0pt}{10pt} & & $\mathscr{P}_\mathrm{T}^2+\mathcal{W}_\mathrm{P}^2 \leq 1$ & no\\ \cline{3-4} 
\rule{0pt}{10pt} & & $\mathscr{P}_\mathrm{T}^2+\mathcal{W}_\mathrm{C}^2 \leq 1$ & no\\  \hline 
\multicolumn{4}{c}{\multirow{2}{*}{Partially distinguishable vs. fully distinguishable particles}}\\ \\ 
\rule{0pt}{10pt} \multirow{3}{*}{$\mathcal{D}_\mathrm{T}$}&$\mathscr{V}_\mathrm{T}$& $\mathcal{D}_\mathrm{T}+\mathscr{V}_\mathrm{T} \leq 1$ & yes\\ \cline{2-4}
\rule{0pt}{10pt} & \multirow{2}{*}{$\mathcal{W}_\mathrm{C}\leq \mathcal{W}_\mathrm{P}$}& $\mathcal{D}_\mathrm{T}^2+\mathcal{W}_\mathrm{P}^2\leq 1$& n/a \\ \cline{3-4}
\rule{0pt}{10pt} & & $\mathcal{D}_\mathrm{V}^2+\mathcal{W}_\mathrm{C}^2\leq 1$& n/a \\
\hline  
\rule{0pt}{10pt} \multirow{3}{*}{$\mathcal{D}_\mathrm{F}$}&$\mathscr{V}_\mathrm{F}$& $\mathcal{D}_\mathrm{F}+\mathscr{V}_\mathrm{F} \leq 1$ & yes\\ \cline{2-4}
\rule{0pt}{10pt} & \multirow{2}{*}{$\mathcal{W}_\mathrm{C}\leq \mathcal{W}_\mathrm{P}$}& $\mathcal{D}_\mathrm{F}^2+\mathcal{W}_\mathrm{P}^2\leq 1$& n/a \\ \cline{3-4}
\rule{0pt}{10pt} & & $\mathcal{D}_\mathrm{F}^2+\mathcal{W}_\mathrm{C}^2\leq 1$& n/a \\
\hline  \\[-8pt] \cline{1-3}
\multicolumn{4}{c}{\multirow{2}{*}{Partially distinguishable vs. indistinguishable particles}}\\ \\ 
\multicolumn{3}{c}{ $D(P^{\mathrm{B}(\mathrm{F})}_\mathcal{M},P_\mathcal{M})+F^2(\rho_\mathrm{E}^{\mathrm{B}(\mathrm{F})},\rho_\mathrm{E}) \leq 1$}& yes
\end{tabular}
\caption{Summary of the relations obtained by comparing the output statistics for initially permuted input particles, between partially distinguishable and fully distinguishable or indistinguishable particles. }
\label{tab:vis}
\end{ruledtabular}
\end{table}

\subsection{Examples for the interference visibility measures}
\label{sec:examples}
\begin{figure}[t]
\centering
\includegraphics[width=\linewidth]{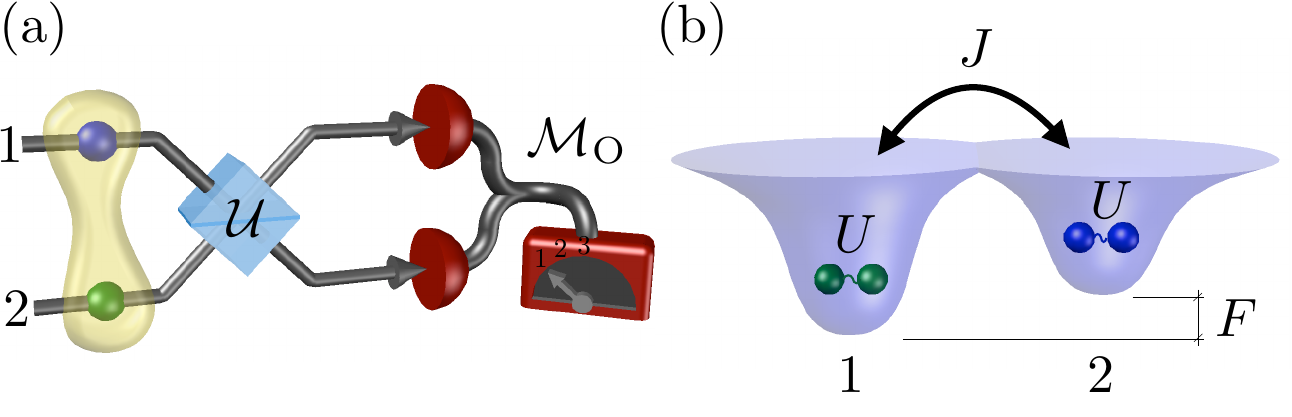}
\caption{Experimental settings to illustrate the complementarity relations~\eqref{eq:Vis02-T} and~\eqref{eq:Vis02}, for systems of partially distinguishable, possibly interacting particles. (a) The Hong-Ou-Mandel experiment: $N=2$ non-interacting partially distinguishable particles (colored balls), which can be correlated in their internal degrees of freedom (yellow envelope), evolve according to $\mathcal{U}=u^{\otimes N}$, with $u$ the single-particle unitary matrix of the balanced beam splitter. The output mode occupation $\vec{S}$ is measured according to the measurement $\mathcal{M}_\mathrm{O}$ (see main text) with an interference contrast controlled by~\eqref{eq:Vis02-T}. (b) The double-well Bose-Hubbard model: Each site (or mode) contains initially two particles, with particles on distinct sites prepared in distinct internal states with mutual overlap $\gamma$ [see main text]. The difference in the potential wells' on-site energies is controlled by the tilt $F$. Particles can tunnel with rate $J$, and interact with strength $U$. After an evolution for some time $t$, different observables exhibit an interference contrast, which is bounded by the complementarity relation~\eqref{eq:Vis02-T}, see Fig.~\ref{fig:Int02}. }
\label{fig:HOMExperiment}
\end{figure}

To conclude, let us illustrate the behaviour of the visibilities $\mathscr{V}_\mathrm{T}$ and $\mathscr{V}_\mathrm{F}$ [see Eqs.~\eqref{eq:Vc} and~\eqref{eq:VcT}] and the inequalities~\eqref{eq:Vis02-T} and \eqref{eq:Vis02} in two experimental scenarios, the Hong-Ou-Mandel experiment, and the double-well Bose-Hubbard model with four partially distinguishable, interacting particles. First we consider the Hong-Ou-Mandel experiment~\cite{Hong-MS-1987,Michler-IB-1996,Mattle-DC-1996} illustrated in Fig.~\ref{fig:HOMExperiment}(a), where $N=2$ non-interacting, partially distinguishable particles (bosons or fermions), are incident on two different input modes of a balanced beam splitter, and measured via a projective measurement of the number of particles in the output modes. For the input state we have $\vec{R}=(1,1)$, $R=2$, $\vec{E}=(1,2)$, and $\Sigma=\mathrm{S}_2=\{\epsilon, (12)\}$, with $\epsilon$ the identity permutation and $(12)$ permuting the particles. In the external basis $\{\ket{\vec{\mathcal{E}}}\}_{\vec{\mathcal{E}}}=\{\ket{1,1},\ket{1,2},\ket{2,1},\ket{2,2}\}$,
the reduced external state~\eqref{eq:rhoE} reads
\begin{align}\label{eq:rhoEExampleHOM}
\rho_\mathrm{E}=\frac{1}{2}\begin{pmatrix}
0&0&0&0\\
0 &1 & r & 0 \\
0 &r & 1 & 0 \\
0&0&0&0
\end{pmatrix},
\end{align}
where the second and third rows and columns correspond to the subset of states $\{\ket{\vec{E}_\mu}\}_{\mu\in\Sigma}=\{\ket{1,2},\ket{2,1}\}$ needed in~\eqref{eq:PsiR} to describe states~\eqref{eq:rho0} with particles in different modes.
For two particles, we find that the non-zero off-diagonal element~\eqref{eq:rhoEoff-diag} is always real, and write $2[\rho_\mathrm{E}]_{\epsilon,(12)}=(-1)_\mathrm{B(F)}\sum_j q_j \bracket{\Omega_{(12)}^{(j)}}{\Omega_\epsilon^{(j)}}=r\in\mathbb{R}$.

The non-interacting evolution of state~\eqref{eq:rhoEExampleHOM} is governed by the unitary $\mathcal{U}=u^{\otimes 2}$, with
\begin{align}\label{eq:uHOM}
u=\frac{1}{\sqrt{2}}\begin{pmatrix}1 & 1\\ 1 & -1\end{pmatrix}
\end{align}
the single particle transformation matrix of the beam splitter. Thereupon, the number of particles in the output modes is measured by $\mathcal{M}_\mathrm{O}=\{M_{\vec{S}}\}_{\vec{S}}$, with $\vec{S}$ the output mode occupation defined analogously to $\vec{R}$ in Sec.~\ref{sec:pd-particles-intro}, and $M_{\vec{S}}=\sum_{\mu\in\Sigma(\vec{S})}\ket{\vec{E}_\mu(\vec{S})}\bra{\vec{E}_\mu(\vec{S})}$ the projector on external states with occupation $\vec{S}$. In total, there are three different output mode occupations $\vec{S}$, which occur with probability $p(\vec{S})=\tr{M_{\vec{S}} \mathcal{U}\rho_\mathrm{E} \mathcal{U}^\dagger}$. These probabilities form the distribution $P_\mathcal{M}$, and read
\begin{align}\label{eq:Phom} 
\begin{split}
p((1,1))=&\frac{1}{2}\left( 1-r  \right),\\
p((2,0))=&p((0,2))=\frac{1}{4}\left( 1+r  \right).
\end{split}
\end{align}
In the case of distinguishable particles, $r=0$, such that
\begin{align*}
\rho_\mathrm{E}^\mathrm{D}=\frac{1}{2}\mathrm{diag}(0,1,1,0),
\end{align*}
and the output probability distribution $P_\mathcal{M}^\mathrm{D}$ becomes
\begin{align}\label{eq:PhomDist}
\begin{split}
p^\mathrm{D}((1,1))=&\frac{1}{2},\\
p^\mathrm{D}((2,0))=&p^\mathrm{D}((0,2))=\frac{1}{4}.
\end{split}
\end{align}

Therewith, a short calculation reveals the distinguishability measures in~\eqref{eq:VisT} and~\eqref{eq:Dc},
\begin{align}\label{eq:DTHOM}
\mathcal{D}_\mathrm{T}=1-\abs{r}, 
\end{align}
and
\begin{align}\label{eq:DVHOM}
\mathcal{D}_\mathrm{F}=\sqrt{1-r^2}. 
\end{align}
Note that in this case, the measures~\eqref{eq:Cmeasure} and~\eqref{eq:Pmeasure} of the many-body state's wave character obey $\mathcal{W}_\mathrm{C}=\mathcal{W}_\mathrm{P}=\abs{r}$, such that $\mathcal{D}_\mathrm{F}^2+\mathcal{W}^2=1$ for $\mathcal{W}\in\{\mathcal{W}_\mathrm{C},\mathcal{W}_\mathrm{P}\}$ [cf. Eq.~\eqref{eq:Connect}].
The probability distributions $P_\mathcal{M}$ and $P_\mathcal{M}^\mathrm{D}$ yield the visibilities from Eqs.~\eqref{eq:VcT} and~\eqref{eq:Vc},
\begin{align}\label{eq:visHOM}
\mathscr{V}_\mathrm{T}= \abs{r} ,
\end{align}
and
\begin{align}\label{eq:visHOMf}
\mathscr{V}_\mathrm{F}=1-\sqrt{ 1-r^2  }.
\end{align}
Interestingly, by virtue of~\eqref{eq:Phom} and~\eqref{eq:PhomDist}, the visibility measure $\mathscr{V}_\mathrm{T}$ from Eq.~\eqref{eq:visHOM} coincides with the usual interference contrast,
\begin{align*}
\mathscr{V}_\mathrm{T}=\abs{ \frac{p^\mathrm{D}(\vec{S})-p(\vec{S})}{p^\mathrm{D}(\vec{S})} }
\end{align*}
for all output events $\vec{S}$. With $\mathcal{D}_\mathrm{T}$ and $\mathscr{V}_\mathrm{T}$ from Eqs.~\eqref{eq:DTHOM} and~\eqref{eq:visHOM} as well as $\mathcal{D}_\mathrm{F}$ and $\mathscr{V}_\mathrm{F}$ from Eqs.~\eqref{eq:DVHOM} and~\eqref{eq:visHOMf}, both inequalities~\eqref{eq:Vis02-T} and~\eqref{eq:Vis02} saturate. Finally, with $\mathcal{W}_\mathrm{C}=\mathcal{W}_\mathrm{P}=\abs{r}$ and Eqs.~\eqref{eq:visHOM} and~\eqref{eq:visHOMf}, we have $\mathscr{V}_\mathrm{F} \leq \mathscr{V}_\mathrm{T} = \mathcal{W}_\mathrm{C}=\mathcal{W}_\mathrm{P}$. With Eq.~\eqref{eq:CompManyAll} this leads to
\begin{align*}
\mathcal{P}^2+\mathscr{V}^2 \leq  1,
\end{align*}
for $\mathcal{P}\in\{\mathcal{P}_\mathrm{T},\mathcal{P}_\mathrm{F}\}$ and $\mathscr{V}\in\{\mathscr{V}_\mathrm{T},\mathscr{V}_\mathrm{F}\}$, in direct analogy with the wave-particle duality relations~\eqref{eq:Englert2} of the double-slit experiment.

Note that for internal product states of bosons (resp. fermions),
\begin{align*}
\ket{\Omega_\epsilon^{(j)}}=\ket{\phi_1^{(j)}} \otimes \ket{\phi_2^{(j)}},
\end{align*}
the off-diagonal element
\begin{align*}
[\rho_\mathrm{E}]_{\epsilon,(12)}= (-1)_\mathrm{B(F)} \frac{1}{2}\sum_j q_j \abs{\bracket{\phi_1^{(j)}}{\phi_2^{(j)}} }^2
\end{align*}
is positive (resp. negative) such that $r\geq 0$ (resp. $r\leq 0$), and the probability of the output event $\vec{S}=(1,1)$ in~\eqref{eq:Phom} decreases (resp. increases) with increasing particle indistinguishability. This refers to the usual scenario of the Hong-Ou-Mandel experiment with uncorrelated bosons (resp. fermions). However, if the particles are in an entangled internal state, this relation can be modified, or even inverted \cite{Michler-IB-1996,Mattle-DC-1996}.

The second example involves the Bose-Hubbard model for partially distinguishable and possibly interacting bosons as illustrated in Fig.~\ref{fig:HOMExperiment}(b). Since the particles' evolution is supposed to be independent of their internal states, we can consider the Bose-Hubbard Hamiltonian with respect to the particles' external degrees of freedom, reading
\begin{align}\label{eq:BHhamil}
H=H_\mathrm{hop}+H_\mathrm{int}+H_\mathrm{tilt}.
\end{align}
In first quantization, the coupling between neighboring sites $\braket{j,k}$ with strength $J$ is described by the hopping term
\begin{align*}
H_\mathrm{hop}=-J \sum_{\braket{j,k}} \sum_{\alpha=1}^N \unit_1 \otimes \dots \otimes \ket{j}\bra{k}_\alpha \otimes \dots \otimes \unit_N,
\end{align*}
where $\ket{j}\bra{k}_\alpha$ acts only on the $\alpha$th particle (and $\unit$ on all other particles). On-site particle interactions of strength $U$ are modeled by
\begin{align*}
H_\mathrm{int}= U \sum_{j=1}^n\sum_{\substack{\alpha,\beta=1 \\\alpha<\beta}}^N  & \unit_1  \otimes  \dots \otimes \ket{j}\bra{j}_\alpha \otimes \dots \\ & \otimes \ket{j}\bra{j}_\beta \otimes \dots \otimes \unit_N,
\end{align*}
and
\begin{align*}
H_\mathrm{tilt}=\sum_{j=1}^n \omega_j \sum_{\alpha=1}^N \unit_1 \otimes \dots \otimes \ket{j}\bra{j}_\alpha \otimes \dots \otimes \unit_N
\end{align*}
additionally accounts for different on-site energies $\omega_j$. Note that in the second quantization formalism these Hamiltonians take their usual form $H_\mathrm{hop}=-J\sum_{\braket{j,k}} a^\dagger_j a_k$, $H_\mathrm{int}=U/2 \sum_{j=1}^n (a^\dagger_j)^2(a_j)^2$, and $H_\mathrm{tilt}=\sum_{j=1}^n \omega_j a^\dagger_j a_j$, with $a^\dagger_j$ (resp. $a_j$) the creation (resp. annihilation) operator of a particle in mode $j$.

In our example, we consider $N=4$ particles in a double-well potential (i.e. $n=2$), where different on-site energies lead to a tilt $F=\omega_2-\omega_1$ [cf. Fig.~\ref{fig:HOMExperiment}(b)]. Initially we consider two particles in each mode, such that $\vec{R}=(2,2)$, $R=6$, $\vec{E}=(1,1,2,2)$, and $\mathrm{S}_{\vec{R}}=\{\epsilon, (12),(34),(12)(34)\}$. Moreover, let us suppose that the particles in mode $1$ share the same internal state $\ket{a}$, while the two particles in mode $2$ are both in state $\gamma \ket{a} + \sqrt{1-\gamma^2} \ket{b}$, with $\gamma\in[0,1]$ and $\bracket{a}{b}=0$. Thus, for increasing $\gamma$, particles in different modes become increasingly indistinguishable. This allows us to study the quantum-to-classical transition in this setting merely with respect to the coefficient $\gamma$ \cite{Ra-NQ-2013}. The many-particle internal state~\eqref{eq:Omega} then reads
\begin{align}\label{eq:OmegaHOM4} 
\begin{split}
\ket{\Omega}= &\gamma^2 \ket{a,a,a,a} + \gamma \sqrt{1-\gamma^2} \ket{a,a,b,a}\\
+& \gamma \sqrt{1-\gamma^2} \ket{a,a,a,b} + (1-\gamma^2) \ket{a,a,b,b},
\end{split}
\end{align}
with the coefficients $C_{\vec{\mathcal{I}}}$ from Eq.~\eqref{eq:Omega} given by $C_{(a,a,a,a)}=\gamma^2$, $C_{(a,a,b,a)}=C_{(a,a,a,b)}=\gamma\sqrt{1-\gamma^2}$, and $C_{(a,a,b,b)}=1-\gamma^2$. As required, this set of coefficients is normalized, $\sum_{\vec{\mathcal{I}}} |C_{\vec{\mathcal{I}}}|^2=1$, and symmetric under the exchange of particles in the same mode, $C_{\vec{\mathcal{I}}}=C_{\vec{\mathcal{I_\xi}}}$ for all $\xi \in \mathrm{S}_{\vec{R}}$ [see below Eq.~\eqref{eq:Omegamu}]. Accordingly, the internal state~\eqref{eq:OmegaHOM4} is normalized, and symmetric, i.e. $\ket{\Omega_\xi}=\ket{\Omega}$ for all $\xi \in \mathrm{S}_{\vec{R}}$.

After the evolution of the many-particle state according to $\mathcal{U}=e^{-iHt/\hbar}$ with $H$ from Eq.~\eqref{eq:BHhamil}, we consider different measurements of the resulting outcome: the projective measurement $\mathcal{M}_\mathrm{O}=\{M_{\vec{S}}\}_{\vec{S}}$ of the output mode occupations $\vec{S}\in\{ (2,2),(3,1),(1,3),(4,0),(0,4)\}$ [see below Eq.~\eqref{eq:uHOM}], the one-point density measurement $\mathcal{M}_\mathrm{1P}=\{M_1,M_2\}$, with 
\begin{align*}
M_j= \frac{1}{N} \sum_{\alpha=1}^N  \unit_1 \otimes \dots \otimes \ket{j}\bra{j}_\alpha \otimes \dots \otimes \unit_N
\end{align*}
measuring the particle density on site $j$, the two-point (density-density) correlations measurement $\mathcal{M}_\mathrm{2P}=\{M_1^2,M_2^2,2M_1 M_2\}$, which measures density correlations between the two sites, as well as three-point and four-point density correlation measurements $\mathcal{M}_\mathrm{3P}=\{M_1^3,M_2^3,3M_1 M_2^2,3M_1^2 M_2\}$, and $\mathcal{M}_\mathrm{4P}=\{M_1^4,M_2^4,4M_1 M_2^3,4M_1^3 M_2,6 M_1^2 M_2^2\}$. As required, these measurements constitute POVMs satisfying $\sum_{M\in\mathcal{M}} M = \unit$.

\begin{figure}[t]
\centering
\includegraphics[width=\linewidth]{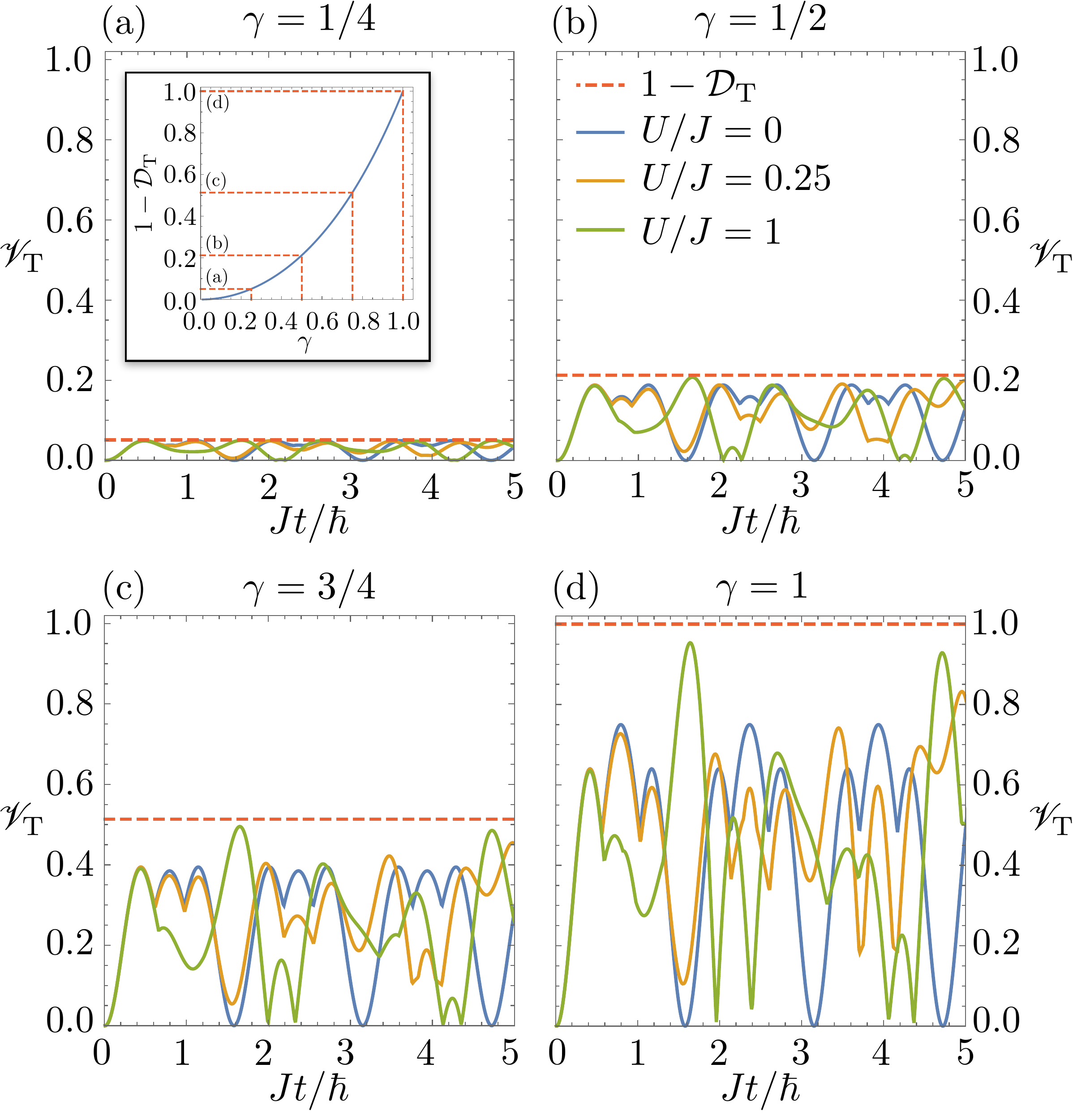}
\caption{Many-particle interference visibility $\mathcal{V}_\mathrm{T}$ in the double-well Bose-Hubbard model~\eqref{eq:BHhamil} for the particle occupation measurement $\mathcal{M}_\mathrm{O}$ and tilt $F=0$. Panels (a)--(d) show the visibility $\mathcal{V}_\mathrm{T}$ as a function of the evolution time $t$ for different levels of particle distinguishability in the internal state~\eqref{eq:OmegaHOM4}, as controlled by $\gamma=1/4$, $1/2$, $3/4$, and $1$, respectively. \emph{Independently} of the interaction strength $U/J$, and as predicted by~\eqref{eq:VisT}, the visibility $\mathcal{V}_\mathrm{T}$ is bounded for all evolution times $t$ by the particle indistinguishability $1-\mathcal{D}_\mathrm{T}$ (red dashed line), which is a monotonously increasing function of $\gamma$ as illustrated in the inset in panel (a). }
\label{fig:Int}
\end{figure}

\begin{figure}[t]
\centering
\includegraphics[width=\linewidth]{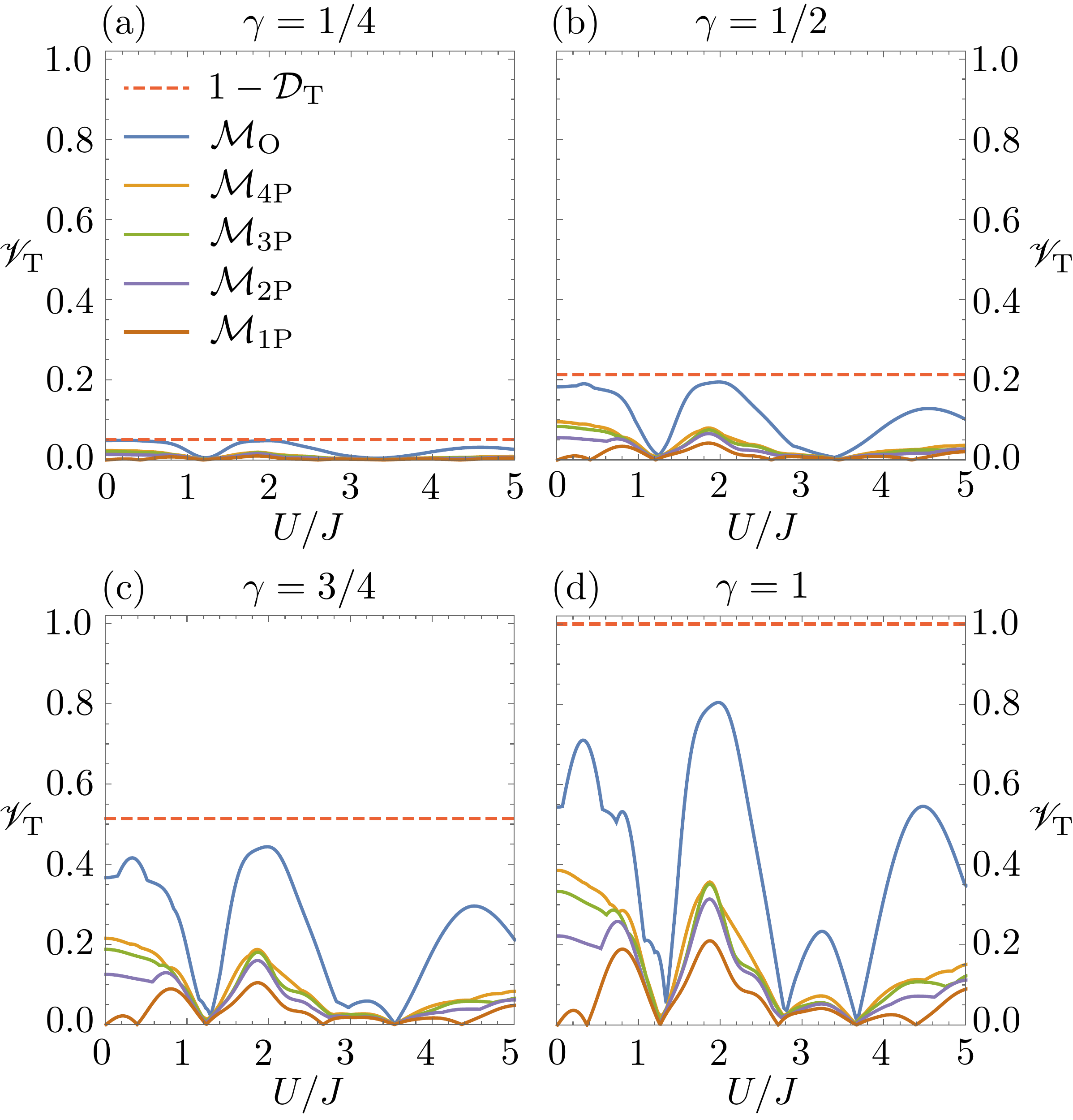}
\caption{Many-particle interference visibility $\mathcal{V}_\mathrm{T}$ in the double-well Bose-Hubbard model for the evolution time $t=4\hbar/J$ and tilt $F=J$. For the particle occupation measurement $\mathcal{M}_\mathrm{O}$, as well as the correlation measurements $\mathcal{M}_\mathrm{1P}$, $\mathcal{M}_\mathrm{2P}$, $\mathcal{M}_\mathrm{3P}$, and $\mathcal{M}_\mathrm{4P}$, panels (a)--(d) show the visibility $\mathcal{V}_\mathrm{T}$ as a function of the interaction strength $U/J$ for $\gamma=1/4$, $1/2$, $3/4$, and $1$, respectively. Just like in Fig.~\ref{fig:Int}, for all measurements and all interaction strengths $U/J$, the visibility $\mathcal{V}_\mathrm{T}$ is bounded by the particle indistinguishability $1-\mathcal{D}_\mathrm{T}$ (red dashed line). See the inset in Fig.~\ref{fig:Int}(a) for the relation between $1-\mathcal{D}_\mathrm{T}$ and $\gamma$.}
\label{fig:Int02}
\end{figure}

First, let us consider the visibility $\mathscr{V}_\mathrm{T}$ from Eq.~\eqref{eq:VcT} for the measurement $\mathcal{M}_\mathrm{O}$ of the output mode occupations, given equal on-site energies, i.e. $F=0$. As highlighted in Figs.~\ref{fig:Int}(a)--(d), the upper bound from Eq.~\eqref{eq:VisT} limits the extent of the interference visibility throughout the evolution (i.e. for all evolution times $t$), and independently of the interaction strength (i.e. for all values of $U/J$). The bound of $\mathscr{V}_\mathrm{T}$ merely depends on particle distinguishability as governed by the overlap $\gamma$ between particles initially in different modes [see the inset in Fig.~\ref{fig:Int}(a)]. In the case of a fixed evolution time $t=4\hbar/J$ and tilt $F=J$, Figs.~\ref{fig:Int02}(a)--(d) show that our bound~\eqref{eq:VisT} is likewise valid for the density correlation measurements $\mathcal{M}_\mathrm{1P}$, $\mathcal{M}_\mathrm{2P}$, $\mathcal{M}_\mathrm{3P}$, and $\mathcal{M}_\mathrm{4P}$. Note that the gap between the visibility $\mathscr{V}_\mathrm{T}$ and the upper bound from Eq.~\eqref{eq:VisT}, e.g. in Fig.~\ref{fig:Int02}(d), is an artefact of the particular measurement. Indeed, by optimizing the measurement [see below Eq.~\eqref{eq:Dc}] the upper bound can be reached. 

In summary, while the exact behavior of the many-particle interference visibility $\mathscr{V}_\mathrm{T}$ undergoes complex dynamics, Figs.~\ref{fig:Int} and~\ref{fig:Int02} illustrate that $\mathscr{V}_\mathrm{T}$ is bounded by particle distinguishability via Eq.~\eqref{eq:VisT}, which applies independently of the particular underlying experimental setting, even beyond the Bose-Hubbard model considered in the present example. Note that the visibility $\mathscr{V}_\mathrm{F}$ from Eq.~\eqref{eq:Vc} obeys a similar behavior.

\section{Conclusion}
\label{sec:conclusion}
Many-particle interference is an intricate phenomenon which leads to complex dynamics for both interacting and non-interacting systems. It is a central ingredient for applications in quantum information processing and quantum simulation, ranging from boson sampling \cite{Aaronson-CC-2013} to simulations of quantum many-body problems with ultracold atoms in optical lattices \cite{Gross-QS-2017}, quantum walks with strongly correlated quantum matter \cite{Preiss-SC-2015}, or relaxation phenomena in a many-body quantum system \cite{Kaufman-QT-2016}.
This calls for a better understanding of the impact of partial particle distinguishability of the involved particles -- which deteriorates any interference-based many-body quantum protocol.

Here, we investigated particle distinguishability in an approach that relies on the complementarity of wave- and particle-like features of many-body quantum states. By deriving the complementarity relations~\eqref{eq:CompManyAll}, we have shown that the fundamental concept of wave-particle duality can be generalized to the many-body realm. We have then demonstrated how the wave and particle properties of a many-body state affect the interference contrast imprinted on diverse experimental observables. In particular, we have established general visibility measures that can be applied \emph{independently} of the specific experimental setting, and thus provide universal tools to quantify the magnitude of many-particle interference, and to certify particle indistinguishability, as illustrated by numerical examples.

Formally, the effect of particle distinguishability on the discussed many-body interference scenarios bears a profound similarity to decoherence due to interactions with an environment. At the single-particle level, this relation is clear: for example, in the double-slit interference of macromolecules \cite{Hornberger-CD-2003,Arndt-PL-2005}, collisions entangle the particle's state with the environment and lead to which-way information that deteriorates single-particle interference. In the many-particle case, we here assumed that the entanglement of internal and external degrees of freedom is given a priori. The creation of such correlations by controlled or uncontrolled interactions of the many-particle system with environmental degrees of freedom mediates decoherence of many-particle amplitudes, and poses a panoply of interesting questions, which are open for future investigations.

\begin{acknowledgements}
The authors would like to thank Heinz-Peter Breuer, Robert Keil, Alberto Rodr\'{\i}guez, and Valery Shchesnovich for fruitful discussions. G.W. and C.D. acknowledge support by the Austrian Science Fund (FWF project I 2562 and F 7114) and the Canadian Institute for Advanced Research (CIFAR, Quantum Information Science Program). C.D. acknowledges the Austrian Academy of Sciences for a DOC fellowship and the Georg H. Endress foundation for financial support. G.D. is grateful to the Alexander von Humboldt foundation for support.
\end{acknowledgements}

\begin{appendix}
\section{Proof of the single-particle wave-particle duality relations~\eqref{eq:Englert2}}
\label{appsec:doubleslit}
In the following we prove the wave-particle duality relations in Eq.~\eqref{eq:Englert2}. Let us consider the fidelity of the detector states $\rho_\mathrm{d}^{\mathrm{A}}$ and $\rho_\mathrm{d}^{\mathrm{B}}$ [see Eq.~\eqref{eq:rhoD}], and utilize its concavity \cite{Nielsen-QC-2011},
\begin{align}
\nonumber F(\rho_\mathrm{d}^{\mathrm{A}},\rho_\mathrm{d}^{\mathrm{B}})\geq&\sum_j q_j F\left( \ketbra{D_\mathrm{A}^{(j)}}{D_\mathrm{A}^{(j)}}, \ketbra{D_\mathrm{B}^{(j)}}{D_\mathrm{B}^{(j)}}\right)\\
\nonumber =&\sum_j q_j \abs{\bracket{D_\mathrm{B}^{(j)}}{D_\mathrm{A}^{(j)}}}\\
\nonumber \geq&\abs{\sum_j q_j \bracket{D_\mathrm{B}^{(j)}}{D_\mathrm{A}^{(j)}}}\\
\label{eq:ineqV}=&\mathcal{V},
\end{align}
where we identified $\mathcal{V}$ from Eq.~\eqref{eq:V} in the last step. By plugging~\eqref{eq:ineqV} into Eq.~\eqref{eq:DF}, we obtain 
\begin{align*}
\mathcal{P}_\mathrm{f}=\sqrt{1-F^2(\rho_\mathrm{d}^{\mathrm{A}},\rho_\mathrm{d}^{\mathrm{B}})}
\leq\sqrt{1-\mathcal{V}^2},
\end{align*}
which results in the second inequality of Eq.~\eqref{eq:Englert2},
\begin{align*}
\mathcal{P}_\mathrm{f}^2+\mathcal{V}^2 \leq 1.
\end{align*}
The first inequality in Eq.~\eqref{eq:Englert2} follows directly from Eq.~\eqref{eq:DDinequDouble}. Note that both inequalities saturate for pure states.

\section{Proof of the hierarchy~\eqref{eq:hierarchy} of quantifiers of the many-body state's wave character}
\label{appsec:hierarchy}
Before we prove the hierarchy in Eq.~\eqref{eq:hierarchy}, let us rewrite the two measures $\mathcal{W}_\mathrm{C}$ and $\mathcal{W}_\mathrm{P}$ from Eqs.~\eqref{eq:Cmeasure} and \eqref{eq:Pmeasure}. With $\rho_\mathrm{E}$ from Eq.~\eqref{eq:rhoE}, the summands of the normalized coherence~\eqref{eq:Cmeasure} can be expressed as
\begin{align}\label{eq:Appoffdiagonal}
\abs{\bra{\vec{E}_\mu}\rho_\mathrm{E}\ket{\vec{E}_\nu}}=\frac{1}{R} \abs{\sum_j q_j \bracket{\Omega_\nu^{(j)}}{\Omega_\mu^{(j)}}},
\end{align}
and by plugging into Eq.~\eqref{eq:Cmeasure} we obtain
\begin{align}\label{CmeasureApp}
\mathcal{W}_\mathrm{C}=\frac{1}{R(R-1)} \sum_{\substack{\mu,\nu\in\Sigma\\ \mu\neq \nu}} \abs{\sum_j q_j \bracket{\Omega_\nu^{(j)}}{\Omega_\mu^{(j)}}}.
\end{align}
Going on to the purity-based measure, a straightforward calculation gives the purity of $\rho_\mathrm{E}$,
\begin{align*}
\tr{\rho_\mathrm{E}^2}=\frac{1}{R^2} \sum_{\mu,\nu\in\Sigma} \abs{\sum_j q_j \bracket{\Omega_\nu^{(j)}}{\Omega_\mu^{(j)}}}^2.
\end{align*}
Since the summands for $\mu = \nu$ yield $1/R$, the normalized purity~\eqref{eq:Pmeasure} becomes
\begin{align}\label{eq:PmeasureApp}
\mathcal{W}_\mathrm{P}=\sqrt{ \frac{1}{R(R-1)} \sum_{\substack{\mu,\nu\in\Sigma\\ \mu\neq \nu}} \abs{\sum_j q_j \bracket{\Omega_\nu^{(j)}}{\Omega_\mu^{(j)}}}^2 }.
\end{align}

Now we are set to prove the hierarchy~\eqref{eq:hierarchy}. We start from Eq.~\eqref{CmeasureApp} and use the Cauchy-Schwarz inequality, 
\begin{align}
\mathcal{W}_\mathrm{C}^2=&\left( \sum_{\substack{\mu,\nu\in\Sigma\\ \mu\neq \nu}} \frac{1}{R(R-1)}  \abs{\sum_j q_j \bracket{\Omega_\nu^{(j)}}{\Omega_\mu^{(j)}}} \right)^2\nonumber \\ 
\leq & \sum_{\substack{\kappa,\tau\in\Sigma\\ \kappa\neq \tau}} \left(\frac{1}{R(R-1)} \right)^2  \sum_{\substack{\mu,\nu\in\Sigma\\ \mu\neq \nu}}\abs{\sum_j q_j \bracket{\Omega_\nu^{(j)}}{\Omega_\mu^{(j)}}}^2\label{eq:WCWPCSinequ}\\
=&\frac{1}{R(R-1)}   \sum_{\substack{\mu,\nu\in\Sigma\\ \mu\neq \nu}}\abs{\sum_j q_j \bracket{\Omega_\nu^{(j)}}{\Omega_\mu^{(j)}}}^2\nonumber \\
=&\mathcal{W}_\mathrm{P}^2,\nonumber
\end{align}
where we identified the square of Eq.~\eqref{eq:PmeasureApp} in the last step. Accordingly,
\begin{align}\label{eq:WcWpApp}
\mathcal{W}_\mathrm{C}\leq \mathcal{W}_\mathrm{P},
\end{align}
which completes the proof. Note that by the use of the Cauchy-Schwarz inequality in~\eqref{eq:WCWPCSinequ}, Eq.~\eqref{eq:WcWpApp} saturates if and only if $ \abs{\sum_j q_j \bracket{\Omega_\nu^{(j)}}{\Omega_\mu^{(j)}}}$ is equal for all $\mu\neq\nu$. Considering Eq.~\eqref{eq:rhoEoff-diag}, this refers to all off-diagonal elements of $\rho_\mathrm{E}$ having equal modulus.

\section{Proof of the hierarchy~\eqref{eq:DDinequality} of quantifiers of the many-body state's particle character}
\label{appsec:DDinequality}
In order to prove Eq.~\eqref{eq:DDinequality} we consider the squared particle character measure $\mathcal{P}_\mathrm{T}$ from Eq.~\eqref{eq:DTT} and use the Fuchs-van de Graaf inequality~\eqref{eq:FvdG}, resulting in
\begin{align*}
\mathcal{P}_\mathrm{T}^2=&\left( \frac{1}{R(R-1)} \sum_{\substack{\mu,\nu\in\Sigma\\ \mu\neq\nu}} D(\rho_\mathrm{I}^{\mu},\rho_\mathrm{I}^{\nu}) \right)^2\\
\leq&\left( \frac{1}{R(R-1)} \sum_{\substack{\mu,\nu\in\Sigma\\ \mu\neq\nu}} \sqrt{1-F^2(\rho_\mathrm{I}^{\mu},\rho_\mathrm{I}^{\nu})} \right)^2.
\end{align*}
Now, using the Cauchy-Schwarz inequality, we arrive at
\begin{align*}
\mathcal{P}_\mathrm{T}^2 \leq& \sum_{\substack{\kappa,\tau\in\Sigma\\ \kappa\neq\tau}} \left(\frac{1}{R(R-1)} \right)^2  \sum_{\substack{\mu,\nu\in\Sigma\\ \mu\neq\nu}} \left(1-F^2(\rho_\mathrm{I}^{\mu},\rho_\mathrm{I}^{\nu})\right)\\
=& \frac{1}{R(R-1)} \sum_{\substack{\mu,\nu\in\Sigma\\ \mu\neq\nu}} \left(1-F^2(\rho_\mathrm{I}^{\mu},\rho_\mathrm{I}^{\nu})\right)\\
=&1- \frac{1}{R(R-1)} \sum_{\substack{\mu,\nu\in\Sigma\\ \mu\neq\nu}} F^2(\rho_\mathrm{I}^{\mu},\rho_\mathrm{I}^{\nu})\\
=&1-\mathcal{F}^2,
\end{align*}
where we used Eq.~\eqref{eq:Ffidel} in the last step. By definition~\eqref{eq:DFF}, this leads to $\mathcal{P}_\mathrm{T}\leq \mathcal{P}_\mathrm{F}$, which was our initial claim.

\section{Proof of the many-particle wave-particle duality relation~\eqref{eq:CompMany}}
\label{appsec:CompProof}
We prove the wave-particle duality relation~\eqref{eq:CompMany} in a similar way as in the proof of $\mathcal{P}^2_\mathrm{f}+\mathcal{V}^2\leq 1$ in Appendix~\ref{appsec:doubleslit}. First of all, we consider the pairwise fidelities of internal states~\eqref{eq:rhoImu} corresponding to different particle labelings and utilize the fidelity's concavity \cite{Nielsen-QC-2011}:
\begin{align}
\nonumber F(\rho_\mathrm{I}^{\mu},\rho_\mathrm{I}^{\nu})\geq&\sum_j q_j F\left( \ketbra{\Omega_\mu^{(j)}}{\Omega_\mu^{(j)}}, \ketbra{\Omega_\nu^{(j)}}{\Omega_\nu^{(j)}}\right)\\
\nonumber =&\sum_j q_j \abs{\bracket{\Omega_\nu^{(j)}}{\Omega_\mu^{(j)}}}\\
\label{eq:FrhoImuconcav}\geq&\abs{\sum_j q_j \bracket{\Omega_\nu^{(j)}}{\Omega_\mu^{(j)}}},
\end{align}
with the inequality saturating for pure internal sates. Therefore, the fidelity measure $\mathcal{F}$ in Eq.~\eqref{eq:Ffidel} obeys
\begin{align*}
\mathcal{F}\geq&\sqrt{  \frac{1}{R(R-1)} \sum_{\substack{\mu,\nu\in\Sigma\\ \mu\neq\nu}} \abs{\sum_j q_j \bracket{\Omega_\nu^{(j)}}{\Omega_\mu^{(j)}}}^2 }\\
=&\mathcal{W}_\mathrm{P},
\end{align*}
where we identified $\mathcal{W}_\mathrm{P}$ from Eq.~\eqref{eq:PmeasureApp}. By means of the definition~\eqref{eq:DFF}, $\mathcal{P}_\mathrm{F}=\sqrt{1-\mathcal{F}^2}$, we then obtain
\begin{align*}
\mathcal{P}_\mathrm{F}\leq\sqrt{1-\mathcal{W}_\mathrm{P}^2}
\end{align*}
and, hence, $\mathcal{P}_\mathrm{F}^2+\mathcal{W}_\mathrm{P}^2 \leq 1$, which had to be proven.

\section{Proof of Eqs.~\eqref{eq:Dequiv} and~\eqref{eq:Fequiv}}
\label{appsec:Permuted}
In order to prove Eqs.~\eqref{eq:Dequiv} and~\eqref{eq:Fequiv}, we first show that $\rho^{\kappa}$ from Eq.~\eqref{eq:rhomu} is obtained by transforming the state $\rho_\mathrm{E}^{\mathrm{B(F)}} \otimes \rho_\mathrm{I}^{\kappa}$ [with $\rho_\mathrm{E}^{\mathrm{B(F)}}$ from Eq.~\eqref{eq:rhoEBF} and $\rho_\mathrm{I}^{\kappa}$ from Eq.~\eqref{eq:rhoImu}] according to the unitary transformation 
\begin{align}\label{eq:Vapp}
V=\sum_{\vec{R}}\sum_{\mu\in\Sigma(\vec{R})} \ket{\vec{E}_\mu(\vec{R})}\bra{\vec{E}_\mu(\vec{R})} \otimes \Pi_\mu,
\end{align}
with the permutation operator $\Pi_\mu$ acting on the internal basis states as
\begin{align*}
\Pi_\mu \ket{\vec{\mathcal{I}}}=&\Pi_\mu \ket{\mathcal{I}_1,\dots,\mathcal{I}_N} \\
=&\ket{\mathcal{I}_{\mu(1)},\dots,\mathcal{I}_{\mu(N)}} \\
=&\ket{\vec{\mathcal{I}}_{\mu}},
\end{align*}
such that
\begin{align*}
\Pi_\mu \ket{\vec{\mathcal{I}}_\kappa}=&\Pi_\mu \ket{(\mathcal{I}_\kappa)_1,\dots,(\mathcal{I}_\kappa)_N} \\
=&\ket{(\mathcal{I}_\kappa)_{\mu(1)},\dots,(\mathcal{I}_\kappa)_{\mu(N)}} \\
=&\ket{\mathcal{I}_{\kappa(\mu(1))},\dots,\mathcal{I}_{\kappa(\mu(N))}} \\
=&\ket{\vec{\mathcal{I}}_{\kappa\mu}},
\end{align*}
and, by Eq.~\eqref{eq:Omega},
\begin{align}\label{eq:LambdaOmega}
\Pi_\mu \ket{\Omega_\kappa^{(j)}}=\ket{\Omega_{\kappa\mu}^{(j)}}.
\end{align}
Note that in Eq.~\eqref{eq:Vapp}, we explicitly state the dependence of the mode assignment list $\vec{E}$ and of the subset $\Sigma$ on the mode occupation $\vec{R}$ since we sum over all possible $\vec{R}$. Given that $\sum_{\vec{R}}\sum_{\mu\in\Sigma(\vec{R})} \ket{\vec{E}_\mu(\vec{R})}\bra{\vec{E}_\mu(\vec{R})}=\sum_{\vec{\mathcal{E}}} \ket{\vec{\mathcal{E}}}\bra{\vec{\mathcal{E}}}=\unit$ and $\Pi_\mu^\dagger \Pi_\mu=\unit$, one checks that $V$ is indeed unitary. Utilizing Eqs.~\eqref{eq:rhoEBF},~\eqref{eq:rhoEBFpure},~\eqref{eq:rhoImu},~\eqref{eq:Vapp} and~\eqref{eq:LambdaOmega}, we then obtain for the transformed state
\begin{align}
&\nonumber V \left[\rho_\mathrm{E}^{\mathrm{B(F)}} \otimes \rho_\mathrm{I}^{\kappa} \right] V^\dagger \\
\nonumber =&\frac{1}{R} \sum_{\mu,\nu\in\Sigma}(-1)^{\mu\nu}_\mathrm{B(F)} \ket{\vec{E}_\mu}\bra{\vec{E}_\nu} \otimes \sum_j q_j \Pi_\mu\ket{\Omega_{\kappa}^{(j)}}\bra{\Omega_{\kappa}^{(j)}}\Pi_\nu^\dagger\\
\nonumber =&\frac{1}{R} \sum_{\mu,\nu\in\Sigma} (-1)^{\mu\nu}_\mathrm{B(F)}\ket{\vec{E}_\mu}\bra{\vec{E}_\nu} \otimes \sum_j q_j \ket{\Omega_{\kappa\mu}^{(j)}}\bra{\Omega_{\kappa\nu}^{(j)}}\\
\label{eq:rhomuVV}=&\rho^{\kappa},
\end{align}
where we identified $\rho^{\kappa}$ from Eq.~\eqref{eq:rhomu}. 

We are now set to prove Eqs.~\eqref{eq:Dequiv} and~\eqref{eq:Fequiv}. Let us start with Eq.~\eqref{eq:Dequiv}: We rewrite the trace distance $D(\rho^{\kappa},\rho^{\tau})$ between permuted states using Eq.~\eqref{eq:rhomuVV} and use the invariance of the trace distance under unitary transformations:
\begin{align}
\nonumber &D(\rho^{\kappa},\rho^{\tau})\\
\nonumber =&D\left(V \left[\rho_\mathrm{E}^{\mathrm{B(F)}} \otimes \rho_\mathrm{I}^{\kappa} \right] V^\dagger,V \left[\rho_\mathrm{E}^{\mathrm{B(F)}} \otimes \rho_\mathrm{I}^{\tau} \right] V^\dagger\right)\\
\label{eq:Dstep}=&D\left(\rho_\mathrm{E}^{\mathrm{B(F)}} \otimes \rho_\mathrm{I}^{\kappa} , \rho_\mathrm{E}^{\mathrm{B(F)}} \otimes \rho_\mathrm{I}^{\tau} \right).
\end{align}
Next, we use $D(\alpha\otimes\sigma,\alpha\otimes\rho)=D(\sigma,\rho)$ for density matrices $\sigma,\rho$ and $\alpha$. This turns Eq.~\eqref{eq:Dstep} into
\begin{align*}
D(\rho^{\kappa},\rho^{\tau})=D( \rho_\mathrm{I}^{\kappa} , \rho_\mathrm{I}^{\tau} ),
\end{align*}
which proves Eq.~\eqref{eq:Dequiv}. Equation~\eqref{eq:Fequiv} can be proven similarly: Utilizing Eq.~\eqref{eq:rhomuVV}, the invariance of the fidelity under unitary transformations and the  multiplicativity of fidelities, we obtain 
\begin{align*}
&F(\rho^{\kappa},\rho^{\tau})\\
=&F\left(V \left[\rho_\mathrm{E}^{\mathrm{B(F)}} \otimes \rho_\mathrm{I}^{\kappa} \right] V^\dagger,V \left[\rho_\mathrm{E}^{\mathrm{B(F)}} \otimes \rho_\mathrm{I}^{\tau} \right] V^\dagger\right)\\
=&F(\rho_\mathrm{E}^{\mathrm{B(F)}} \otimes \rho_\mathrm{I}^{\kappa} , \rho_\mathrm{E}^{\mathrm{B(F)}} \otimes \rho_\mathrm{I}^{\tau})\\
=&F(\rho_\mathrm{E}^{\mathrm{B(F)}}, \rho_\mathrm{E}^{\mathrm{B(F)}} ) F(\rho_\mathrm{I}^{\kappa},\rho_\mathrm{I}^{\tau})\\
=&F(\rho_\mathrm{I}^{\kappa},\rho_\mathrm{I}^{\tau}),
\end{align*}
which proves Eq.~\eqref{eq:Fequiv}.

\section{Proof of Eq.~\eqref{eq:Cmunuclass}}
\label{appsec:Cmunu}
The first inequality in Eq.~\eqref{eq:Cmunuclass} is a statement of the Fuchs-van de Graaf inequality \eqref{eq:DistanceIneqClas}, we therefore set out to prove the second inequality.
We start by utilizing the relation~\eqref{eq:relationF} between the Bhattacharyya coefficient and the quantum fidelity, and, thereafter, the invariance of the quantum fidelity under unitary transformations,
\begin{align*}
F(P_\mathcal{M}^\kappa,P_\mathcal{M}^\tau)\geq& \min_{\mathcal{M}} F(P_\mathcal{M}^\kappa,P_\mathcal{M}^\tau)\\
=&F(\mathcal{U} \rho_\mathrm{E}^{\kappa}\mathcal{U}^\dagger, \mathcal{U} \rho_\mathrm{E}^{\tau}\mathcal{U}^\dagger)\\
=&F( \rho_\mathrm{E}^{\kappa},  \rho_\mathrm{E}^{\tau}).
\end{align*}
Since the quantum fidelity increases under partial trace operations \cite{Nielsen-QC-2011},
and using Eq.~\eqref{eq:Fequiv}, we find
\begin{align}
\nonumber F(P_\mathcal{M}^\kappa,P_\mathcal{M}^\tau)\geq&F( \rho_\mathrm{E}^{\kappa},  \rho_\mathrm{E}^{\tau})\\
\nonumber \geq&F( \rho^{\kappa},  \rho^{\tau})\\
\label{eq:FineqFi}=&F( \rho_\mathrm{I}^{\kappa},  \rho_\mathrm{I}^{\tau}),
\end{align}
with $\rho^{\kappa}$ from Eq.~\eqref{eq:rhomu}. Using Eq.~\eqref{eq:rhoImu} and the concavity of the quantum fidelity [see Eq.~\eqref{eq:FrhoImuconcav}] then yields
\begin{align}
\nonumber F(P_\mathcal{M}^\kappa,P_\mathcal{M}^\tau) \geq&F( \rho_\mathrm{I}^{\kappa},  \rho_\mathrm{I}^{\tau})\\
\nonumber \geq& \sum_j q_j |\bracket{\Omega_\tau^{(j)}}{\Omega_\kappa^{(j)}}|\\
\label{eq:Fcalgeq}\geq& \abs{\sum_j q_j \bracket{\Omega_\tau^{(j)}}{\Omega_\kappa^{(j)}}}.
\end{align}
In view of Eq.~\eqref{eq:Appoffdiagonal} we therefore obtain
\begin{align*}
F(P_\mathcal{M}^\kappa,P_\mathcal{M}^\tau) \geq R \abs{\bra{\vec{E}_\kappa}\rho_\mathrm{E}\ket{\vec{E}_\tau}},
\end{align*}
such that
\begin{align*}
\sqrt{1-F^2(P_\mathcal{M}^\kappa,P_\mathcal{M}^\tau)}
\leq  \sqrt{1-R^2 \abs{\bra{\vec{E}_\kappa}\rho_\mathrm{E}\ket{\vec{E}_\tau}}^2},
\end{align*}
which is the second inequality in~\eqref{eq:Cmunuclass}.

\section{Proof of the hierarchy~\eqref{eq:DcalDcal} of quantifiers of the particle character }
\label{appsec:DcalDcal}
Equation~\eqref{eq:DcalDcal} can be proven by following the same route as in  Appendix~\ref{appsec:DDinequality}. Starting from the definition~\eqref{eq:DcalT} and using Eq.~\eqref{eq:DistanceIneqClas}, we obtain 
\begin{align*}
\mathscr{P}_\mathrm{T}^2 =&\left( \frac{1}{R(R-1)} \sum_{\substack{\kappa,\tau\in\Sigma\\ \kappa\neq\tau}} D(P_\mathcal{M}^\kappa,P_\mathcal{M}^\tau) \right)^2\\
\leq& \left( \frac{1}{R(R-1)} \sum_{\substack{\kappa,\tau\in\Sigma\\ \kappa\neq\tau}} \sqrt{1-F^2(P_\mathcal{M}^\kappa,P_\mathcal{M}^\tau)} \right)^2.
\end{align*}
Using the Cauchy-Schwarz inequality then leads to
\begin{align*}
\mathscr{P}_\mathrm{T}^2 \leq& \sum_{\substack{\mu,\nu\in\Sigma\\ \mu\neq\nu}} \left(\frac{1}{R(R-1)} \right)^2  \sum_{\substack{\kappa,\tau\in\Sigma\\ \kappa\neq\tau}} \left[1-F^2(P_\mathcal{M}^\kappa,P_\mathcal{M}^\tau)\right]\\
=& \frac{1}{R(R-1)} \sum_{\substack{\kappa,\tau\in\Sigma\\ \kappa\neq\tau}} \left[1-F^2(P_\mathcal{M}^\kappa,P_\mathcal{M}^\tau)\right]\\
=&1- \frac{1}{R(R-1)} \sum_{\substack{\kappa,\tau\in\Sigma\\ \kappa\neq\tau}} F^2(P_\mathcal{M}^\kappa,P_\mathcal{M}^\tau)\\
=&\mathscr{P}_\mathrm{F}^2,
\end{align*}
where we identified $\mathscr{P}_\mathrm{F}$ from Eq.~\eqref{eq:DcalF} in the last step.

\section{Proof of Eq.~\eqref{eq:DTDFinequ}}
\label{appsec:CompDcal}
We now prove the inequalities in Eq.~\eqref{eq:DTDFinequ}. We start by proving that $\mathscr{P}_\mathrm{T}\leq \mathcal{P}_\mathrm{T}$. Given Eq.~\eqref{eq:DcalT}, let us maximize the Kolmogorov distance and utilize Eq.~\eqref{eq:relationD} as well as the invariance of the trace distance under unitary transformations, 
\begin{align*}
D(P_\mathcal{M}^\kappa, P_\mathcal{M}^\tau) \leq & \max_{\mathcal{M}}D(P_\mathcal{M}^\kappa, P_\mathcal{M}^\tau)\\
=&D(\mathcal{U}\rho_\mathrm{E}^{\kappa}\mathcal{U}^\dagger, \mathcal{U}\rho_\mathrm{E}^{\tau}\mathcal{U}^\dagger)\\
=&D(\rho_\mathrm{E}^{\kappa}, \rho_\mathrm{E}^{\tau}).
\end{align*}
Now, using the contractivity of the trace distance under partial trace \cite{Nielsen-QC-2011} as well as Eq.~\eqref{eq:Dequiv} yields
\begin{align}
\nonumber D(P_\mathcal{M}^\kappa, P_\mathcal{M}^\tau) \leq &D(\rho_\mathrm{E}^{\kappa}, \rho_\mathrm{E}^{\tau})\\
\nonumber \leq &D(\rho^{\kappa}, \rho^{\tau})\\
\label{eq:DleqD}=&D(\rho_\mathrm{I}^{\kappa}, \rho_\mathrm{I}^{\tau}).
\end{align}
By plugging Eq.~\eqref{eq:DleqD} into~\eqref{eq:DcalT} we then obtain
\begin{align*}
\mathscr{P}_\mathrm{T} \leq \frac{1}{R(R-1)} \sum_{\substack{\kappa,\tau\in\Sigma\\ \kappa\neq\tau}} D(\rho_\mathrm{I}^{\kappa}, \rho_\mathrm{I}^{\tau})=\mathcal{P}_\mathrm{T},
\end{align*}
where we identified $\mathcal{P}_\mathrm{T}$ from Eq.~\eqref{eq:DTT} in the last step. In order to prove the second inequality in Eq.~\eqref{eq:DTDFinequ} we plug Eq.~\eqref{eq:FineqFi} in~\eqref{eq:DcalF} such that
\begin{align*}
\mathscr{P}_\mathrm{F} \leq \sqrt{1-  \frac{1}{R(R-1)} \sum_{\substack{\kappa,\tau\in\Sigma\\ \kappa\neq\tau}} F^2(\rho_\mathrm{I}^{\kappa},\rho_\mathrm{I}^{\tau})   }=\mathcal{P}_\mathrm{F}.
\end{align*}
In the last step we identified $\mathcal{P}_\mathrm{F}$ from Eq.~\eqref{eq:DFF} which finishes the proof.

\section{Proof of the eigenvalue equation~\eqref{eq:eigenequation}}
\label{appsec:Eigenequation}
We provide two proofs of the eigenvalue equation~\eqref{eq:eigenequation}, i.e.~that $\ket{\psi_\mathrm{B(F)}}$ from Eq.~\eqref{eq:rhoEBFpure} is an eigenvector of $\rho_\mathrm{E}$ from Eq.~\eqref{eq:rhoE} with eigenvalue $\lambda_\mathrm{B(F)}= F^2(\rho_\mathrm{E}^{\mathrm{B(F)}},\rho_\mathrm{E})$.

\textit{First proof}: Let us consider the squared fidelity $F^2(\rho_\mathrm{E}^{\mathrm{B(F)}},\rho_\mathrm{E})$, and note that both states, $\rho_\mathrm{E}^{\mathrm{B}(\mathrm{F})}$ and $\rho_\mathrm{E}$, are not necessarily of the same particle type, e.g. if $\rho_\mathrm{E}$ describes a state of fermions, then $F^2(\rho_\mathrm{E}^{\mathrm{B}},\rho_\mathrm{E})$ corresponds to the squared fidelity between a bosonic and a fermionic state. Therefore, for $\rho_\mathrm{E}$ we abbreviate the case of bosons (fermions) by $\mathrm{B}'$ ($\mathrm{F}'$), and for the pure state $\rho_\mathrm{E}^{\mathrm{B}(\mathrm{F})}$ by $\mathrm{B}$ ($\mathrm{F}$). Starting from Eqs.~\eqref{eq:rhoE} and~\eqref{eq:rhoEBFpure}, the squared fidelity can be written as
\begin{align}
&F^2(\rho_\mathrm{E}^{\mathrm{B}(\mathrm{F})},\rho_\mathrm{E})\nonumber \\
=&\bra{\psi_{\mathrm{B}(\mathrm{F})}} \rho_\mathrm{E} \ket{\psi_{\mathrm{B}(\mathrm{F})}}\nonumber\\
=&\frac{1}{R^2} \sum_{\mu\in\Sigma} \left( \sum_{\nu\in\Sigma} \sum_j q_j\ (-1)^{\mu\nu}_\mathrm{B(F)}  (-1)^{\mu\nu}_\mathrm{B'(F')}\bracket{\Omega_\nu^{(j)}}{\Omega_\mu^{(j)}}\right). \label{eq:fidelitysq}
\end{align}
Next, consider the expression in parenthesis, for brevity denoted by $Z$, and insert a sum over all $\xi \in \mathrm{S}_{\vec{R}}$,
\begin{align}\label{eq:Z1app}
Z=& \sum_{\nu\in\Sigma} \sum_j q_j\ (-1)^{\mu\nu}_\mathrm{B(F)}  (-1)^{\mu\nu}_\mathrm{B'(F')}\bracket{\Omega_\nu^{(j)}}{\Omega_\mu^{(j)}}\\
 =&\frac{1}{|\mathrm{S}_{\vec{R}}|} \sum_{\xi \in \mathrm{S}_{\vec{R}}} \sum_{\nu\in\Sigma} \sum_j q_j\ (-1)^{\mu\nu}_\mathrm{B(F)}  (-1)^{\mu\nu}_\mathrm{B'(F')}\bracket{\Omega_\nu^{(j)}}{\Omega_\mu^{(j)}}\nonumber  \\
 =&\frac{1}{|\mathrm{S}_{\vec{R}}|} \sum_{\xi \in \mathrm{S}_{\vec{R}}} \sum_{\nu\in\Sigma} \sum_j q_j\ (-1)^{\mu\nu\xi}_\mathrm{B(F)}  (-1)^{\mu\nu\xi}_\mathrm{B'(F')}\bracket{\Omega_{\xi\nu}^{(j)}}{\Omega_\mu^{(j)}}.\nonumber 
\end{align}
In the last step we used $\ket{\Omega_{\xi\nu}^{(j)}}=(-1)^{\xi}_\mathrm{B'(F')} \ket{\Omega_{\nu}^{(j)}}$, which can be seen by considering the coefficients' symmetry $C_{\vec{\mathcal{I}}_\xi}^{(j)}=(-1)_\mathrm{B'(F')}^\xi C_{\vec{\mathcal{I}}}^{(j)} $ in Eq.~\eqref{eq:Omegamusym}. Furthermore, we inserted the factor $(-1)_\mathrm{B(F)}^\xi$. This can be done due to $\rho_\mathrm{E}^\mathrm{B(F)}$ describing a state of indistinguishable bosons or fermions, with $(-1)_\mathrm{B(F)}^\xi=1$ in the case of bosons, and $\mathrm{S}_{\vec{R}}=\{\epsilon\}$, with $\epsilon$ the identity permutation, in the case of fermions. The latter is a consequence of Pauli's principle. Now we use that the permutations $\pi \in \mathrm{S}_N$ can be decomposed as $\pi=\xi \nu$, with $\xi \in \mathrm{S}_{\vec{R}}$ and $\nu \in \Sigma$, such that
\begin{align*}
Z=&\frac{1}{|\mathrm{S}_{\vec{R}}|} \sum_{\pi\in\mathrm{S}_N} \sum_j q_j\ (-1)^{\mu\pi}_\mathrm{B(F)}  (-1)^{\mu\pi}_\mathrm{B'(F')}\bracket{\Omega_{\pi}^{(j)}}{\Omega_\mu^{(j)}}\\
=&\frac{1}{|\mathrm{S}_{\vec{R}}|} \sum_{\pi\in\mathrm{S}_N} \sum_j q_j\ (-1)^{\pi}_\mathrm{B(F)}  (-1)^{\pi}_\mathrm{B'(F')}\bracket{\Omega_{\pi}^{(j)}}{\Omega_\epsilon^{(j)}}.
\end{align*}
Again using the decomposition $\pi=\xi \nu$, with $\xi \in \mathrm{S}_{\vec{R}}$ and $\nu \in \Sigma$, the symmetry $\ket{\Omega_{\xi\nu}^{(j)}}=(-1)^{\xi}_\mathrm{B'(F')} \ket{\Omega_{\nu}^{(j)}}$, as well as the freedom to insert $(-1)_\mathrm{B(F)}^\xi$ yields
\begin{align}
Z=&\frac{1}{|\mathrm{S}_{\vec{R}}|}  \sum_{\xi \in \mathrm{S}_{\vec{R}}} \sum_{\nu\in\Sigma} \sum_j q_j\ (-1)^{\nu}_\mathrm{B(F)}  (-1)^{\nu}_\mathrm{B'(F')}\bracket{\Omega_\nu^{(j)}}{\Omega_\epsilon^{(j)}}\nonumber \\
=&\sum_{\nu\in\Sigma} \sum_j q_j\ (-1)^{\nu}_\mathrm{B(F)}  (-1)^{\nu}_\mathrm{B'(F')}\bracket{\Omega_\nu^{(j)}}{\Omega_\epsilon^{(j)}}. \label{eq:eqZ2app}
\end{align}
Thus, in consideration of Eqs.~\eqref{eq:Z1app} and~\eqref{eq:eqZ2app} we have
\begin{align}\label{eq:epsilonId}
\nonumber Z=&\sum_{\nu \in\Sigma} \sum_j q_j\ (-1)^{\mu\nu}_\mathrm{B(F)} (-1)^{\mu\nu}_\mathrm{B'(F')} \bracket{\Omega_\nu^{(j)}}{\Omega_\mu^{(j)}}\\
=& \sum_{\nu\in\Sigma} \sum_j q_j\ (-1)^{\nu}_\mathrm{B(F)} (-1)^{\nu}_\mathrm{B'(F')}\bracket{\Omega_\nu^{(j)}}{\Omega_\epsilon^{(j)}},
\end{align}
with $\epsilon$ the identity permutation. Therewith, the squared quantum fidelity in~\eqref{eq:fidelitysq} becomes
\begin{align}
&F^2(\rho_\mathrm{E}^{\mathrm{B}(\mathrm{F})},\rho_\mathrm{E})\nonumber \\
=&\frac{1}{R} \sum_{\nu\in\Sigma} \sum_j q_j\ (-1)^{\nu}_\mathrm{B(F)}(-1)^{\nu}_\mathrm{B'(F')} \bracket{\Omega_\nu^{(j)}}{\Omega_\epsilon^{(j)}}.\label{eq:lambdaEp}
\end{align}

Next we rewrite the left-hand side of Eq.~\eqref{eq:eigenequation} with the help of Eqs.~\eqref{eq:rhoE} and~\eqref{eq:rhoEBFpure}:
\begin{align*}
&\rho_\mathrm{E}\ket{\psi_\mathrm{B(F)}}\\
=&\frac{1}{R^{3/2}} \sum_{\mu,\nu,\tau\in\Sigma} (-1)^{\tau}_\mathrm{B(F)}(-1)^{\mu\nu}_\mathrm{B'(F')}\\
&\times \sum_j q_j \bracket{\Omega_\nu^{(j)}}{\Omega_\mu^{(j)}} \ket{\vec{E}_\mu}\bracket{\vec{E}_\nu}{\vec{E}_\tau}\\
=&\frac{1}{R^{3/2}}\sum_{\mu,\nu\in\Sigma}  \sum_j q_j\  (-1)^{\nu}_\mathrm{B(F)}(-1)^{\mu\nu}_\mathrm{B'(F')} \bracket{\Omega_\nu^{(j)}}{\Omega_\mu^{(j)}} \ket{\vec{E}_\mu}.
\end{align*}
Using Eqs.~\eqref{eq:epsilonId} and~\eqref{eq:lambdaEp} then yields
\begin{align*}
&\rho_\mathrm{E}\ket{\psi_\mathrm{B(F)}}\\
=&\frac{1}{R^{3/2}}\sum_{\mu,\nu\in\Sigma}  \sum_j q_j\  (-1)^{\mu\nu}_\mathrm{B(F)}(-1)^{\nu}_\mathrm{B'(F')} \bracket{\Omega_\nu^{(j)}}{\Omega_\epsilon^{(j)}} \ket{\vec{E}_\mu}\\
=&F^2(\rho_\mathrm{E}^{\mathrm{B}(\mathrm{F})},\rho_\mathrm{E}) \frac{1}{\sqrt{R}}\sum_{\mu\in\Sigma} (-1)^{\mu}_\mathrm{B(F)} \ket{\vec{E}_\mu}\\
=&F^2(\rho_\mathrm{E}^{\mathrm{B}(\mathrm{F})},\rho_\mathrm{E})  \ket{\psi_\mathrm{B(F)}},
\end{align*}
where we recognized $\ket{\psi_\mathrm{B(F)}}$ from Eq.~\eqref{eq:rhoEBFpure} in the last step. Thus, we identify the eigenvalue $\lambda_\mathrm{B(F)}=F^2(\rho_\mathrm{E}^{\mathrm{B}(\mathrm{F})},\rho_\mathrm{E}) $, which finishes the proof.

\textit{Second proof}: One arrives at the same result faster by using a result from group representation theory known as unitary-unitary duality \cite{Rowe-DP-2012}. One can indeed show that the reduced external state can be decomposed according to the irreducible representations $\Lambda$ of the symmetric group $\mathrm{S}_N$:
\begin{align}\label{eq:eigendecomp}
\rho_\mathrm{E}=\bigoplus_\Lambda \lambda_\Lambda\ \rho_\Lambda,
\end{align}
In particular, the totally symmetric, $\Lambda=\mathrm{B}$, and totally antisymmetric, $\Lambda=\mathrm{F}$, irreducible representations each contain only one state with mode occupation $\vec{R}$: the bosonic and fermionic states $\rho_\mathrm{B}=\ket{\psi_\mathrm{B}}\bra{\psi_\mathrm{B}}$ and $\rho_\mathrm{F}=\ket{\psi_\mathrm{F}}\bra{\psi_\mathrm{F}}$, respectively, as given by Eq.~\eqref{eq:rhoEBFpure}, which must therefore appear in the decomposition~\eqref{eq:eigendecomp}.
The identity $\lambda_\mathrm{B(F)}= F^2(\rho_\mathrm{E}^{\mathrm{B(F)}},\rho_\mathrm{E})$ then follows from
\begin{align*}
F^2(\rho_\mathrm{E}^{\mathrm{B(F)}},\rho_\mathrm{E})=&\bra{\psi_\mathrm{B(F)}}\rho_\mathrm{E} \ket{\psi_\mathrm{B(F)}}\\
=&\lambda_\mathrm{B(F)}.
\end{align*}

\section{Proof of Eq.~\eqref{eq:FvGspecial}}
\label{appsec:FvGspecial}

The proof of Eq.~\eqref{eq:FvGspecial} is based on Eq.~\eqref{eq:eigenequation}, which allows us to decompose $\rho_\mathrm{E}$ as
\begin{align}\label{eq:rhoEdecomp}
\rho_\mathrm{E}=\lambda_\mathrm{B(F)}\ \rho_\mathrm{E}^{\mathrm{B(F)}} + (1-\lambda_\mathrm{B(F)})\ \rho_\mathrm{E}^{\perp},
\end{align}
where $\rho_\mathrm{E}^{\mathrm{B(F)}}$ and $\rho_\mathrm{E}^{\perp}$ have support on orthogonal subspaces, such that their trace distance yields
\begin{align}\label{eq:Orth}
D(\rho_\mathrm{E}^{\mathrm{B(F)}},\rho_\mathrm{E}^{\perp})=1.
\end{align}
By utilizing Eqs.~\eqref{eq:rhoEdecomp} and~\eqref{eq:Orth} as well as the convexity of the trace distance \cite{Nielsen-QC-2011}, we obtain the upper bound
\begin{align}
\nonumber D(\rho_\mathrm{E}^{\mathrm{B(F)}},\rho_\mathrm{E}) \leq & \lambda_\mathrm{B(F)}\ D(\rho_\mathrm{E}^{\mathrm{B(F)}},\rho_\mathrm{E}^{\mathrm{B(F)}})\\
\nonumber &+ (1-\lambda_\mathrm{B(F)})\ D(\rho_\mathrm{E}^{\mathrm{B(F)}},\rho_\mathrm{E}^{\perp})\\
\nonumber =&1-\lambda_\mathrm{B(F)}\\
\label{eq:DFupper}=&1-F^2(\rho_\mathrm{E}^{\mathrm{B(F)}},\rho_\mathrm{E}),
\end{align}
where we used $\lambda_\mathrm{B(F)}= F^2(\rho_\mathrm{E}^{\mathrm{B(F)}},\rho_\mathrm{E})$ in the last step [see Eq.~\eqref{eq:eigenequation}]. On the other hand, the known lower bound on the trace distance if at least one state is pure \cite{Nielsen-QC-2011}, $D(\ket{\Psi},\rho)\geq 1- F^2(\ket{\Psi},\rho)$, leads to
\begin{align}\label{eq:DFlower}
D(\rho_\mathrm{E}^{\mathrm{B(F)}},\rho_\mathrm{E}) \geq 1-F^2(\rho_\mathrm{E}^{\mathrm{B(F)}},\rho_\mathrm{E}).
\end{align}
Combining Eqs.~\eqref{eq:DFupper} and~\eqref{eq:DFlower} results in
\begin{align*}
D(\rho_\mathrm{E}^{\mathrm{B(F)}},\rho_\mathrm{E}) +F^2(\rho_\mathrm{E}^{\mathrm{B(F)}},\rho_\mathrm{E})= 1,
\end{align*}
which had to be proven.

\section{Proof of Eq.~\eqref{eq:Vis01}}
\label{appsec:Vis01}
We now prove Eq.~\eqref{eq:Vis01}. Utilizing the relation~\eqref{eq:relationD} of the Kolmogorov distance to the trace norm, as well as the invariance of the trace distance under unitary transformations \cite{Nielsen-QC-2011}, we have
\begin{align}
\label{eq:Dmax} D(P^{\mathrm{B}(\mathrm{F})}_\mathcal{M},P_\mathcal{M})\leq& \max_{\mathcal{M}}D(P^{\mathrm{B}(\mathrm{F})}_\mathcal{M},P_\mathcal{M})\\
\nonumber=&D(\mathcal{U}\rho_\mathrm{E}^{\mathrm{B(F)}}\mathcal{U}^\dagger, \mathcal{U}\rho_\mathrm{E}\mathcal{U}^\dagger)\\ \label{eq:Dprove1}
=&D(\rho_\mathrm{E}^{\mathrm{B(F)}}, \rho_\mathrm{E}).
\end{align}
Plugging Eq.~\eqref{eq:Dprove1} into~\eqref{eq:FvGspecial} then leads to
\begin{align}\label{eq:Vis01app}
D(P^{\mathrm{B}(\mathrm{F})}_\mathcal{M},P_\mathcal{M})+F^2(\rho_\mathrm{E}^{\mathrm{B(F)}},\rho_\mathrm{E})\leq 1,
\end{align}
which is the sought-after relation. Note that the inequality in Eq.~\eqref{eq:Vis01app} [and, thus, in Eq.~\eqref{eq:Vis01}] is only due to the maximization in Eq.~\eqref{eq:Dmax} and saturates for an optimal measurement. One such measurement is given by the projection onto the eigenstates of $\mathcal{U}(\rho_\mathrm{E}-\rho_\mathrm{E}^\mathrm{B(F)})\mathcal{U}^\dagger$ [see also Sec.~9.2.1 in \cite{Nielsen-QC-2011}].
\section{Proof of Eq.~\eqref{eq:Connect}}
\label{appsec:Connect}
Let us start with proving Eq.~\eqref{eq:Connect} for the distinguishability measure $\mathcal{D}_\mathrm{T}$ by utilizing the lower bound of the trace distance between density operators $\rho$ and $\sigma$ derived in Ref.~\cite{Puchala-BT-2009} [Theorem~1 there],
\begin{align*}
D(\rho,\sigma) \geq 1- \tr{\rho\sigma}- \sqrt{1-\tr{\rho^2}} \sqrt{1-\tr{\sigma^2}}.
\end{align*}
For the trace distance between $\rho_\mathrm{E}$ and $\rho_\mathrm{E}^\mathrm{D}$ from Eqs.~\eqref{eq:rhoE} and~\eqref{eq:rhoEdist}, we can use $\tr{\rho_\mathrm{E}^\mathrm{D} \rho_\mathrm{E}}=\tr{(\rho_\mathrm{E}^\mathrm{D})^2}=1/R$, such that
\begin{align*}
D(\rho_\mathrm{E}^\mathrm{D},\rho_\mathrm{E}) \geq  \frac{R-1}{R} - \sqrt{ \frac{R-1}{R} \left( 1-\tr{\rho_\mathrm{E}^2}  \right) }.
\end{align*}
Considering Eq.~\eqref{eq:Pmeasure}, we have
\begin{align}\label{eq:1-tr}
1- \tr{\rho_\mathrm{E}^2} = \frac{R-1}{R}\left(1-\mathcal{W}_\mathrm{P}^2\right),
\end{align}
which leads us to
\begin{align*}
D(\rho_\mathrm{E}^\mathrm{D},\rho_\mathrm{E}) \geq  \frac{R-1}{R} \left( 1- \sqrt{ 1-\mathcal{W}_\mathrm{P}^2 } \right).
\end{align*}
By plugging into Eq.~\eqref{eq:VisT} we obtain
\begin{align*}
\mathcal{D}_\mathrm{T} =& 1- \frac{R}{R-1} D(\rho_\mathrm{E}^\mathrm{D},\rho_\mathrm{E}) \\
\leq &\sqrt{ 1-\mathcal{W}_\mathrm{P}^2 },
\end{align*}
and, accordingly, $\mathcal{D}_\mathrm{T}^2+\mathcal{W}_\mathrm{P}^2 \leq 1$. Together with the hierarchy~\eqref{eq:hierarchy}, this proves Eq.~\eqref{eq:Connect} for $\mathcal{D}_\mathrm{T}$.

Next we prove Eq.~\eqref{eq:Connect} for the distinguishability measure $\mathcal{D}_\mathrm{F}$. Therefore, we first of all recall that $\rho_\mathrm{E}^{\mathrm{D}}$ from Eq.~\eqref{eq:rhoEdist} is maximally mixed, such that $\rho_\mathrm{E}^{\mathrm{D}}$ and $\rho_\mathrm{E}$ can be simultaneously diagonalized. Letting $\lambda_\alpha$ be the eigenvalues of $\rho_\mathrm{E}$, the fidelity of
 $\rho_\mathrm{E}^{\mathrm{D}}$ and $\rho_\mathrm{E}$ can thus be written as \cite{Fuchs-CD-1999,Nielsen-QC-2011}
\begin{align}\label{eq:FappD}
F(\rho_\mathrm{E}^{\mathrm{D}},\rho_\mathrm{E})=\sum_{\alpha=1}^R \sqrt{ \frac{1}{R}\lambda_\alpha}.
\end{align}
Taking the square of Eq.~\eqref{eq:FappD} and utilizing $\sum_\alpha \lambda_\alpha=1$ yields
\begin{align*}
F^2(\rho_\mathrm{E}^{\mathrm{D}},\rho_\mathrm{E})=&\sum_{\alpha,\beta=1}^R \frac{1}{R}\sqrt{\lambda_\alpha \lambda_\beta}\\
=&\frac{1}{R} +\sum_{\substack{\alpha,\beta=1 \\ \alpha\neq\beta}}^R \frac{1}{R}\sqrt{\lambda_\alpha \lambda_\beta}.
\end{align*}
Now, using the Cauchy-Schwarz inequality in the second summand, we obtain
\begin{align}
\label{eq:CSuse} F^2(\rho_\mathrm{E}^{\mathrm{D}},\rho_\mathrm{E})\leq&\frac{1}{R} +\sqrt{\sum_{\substack{\gamma,\delta=1 \\ \gamma\neq\delta}}^R \frac{1}{R^2}\sum_{\substack{\alpha,\beta=1 \\ \alpha\neq\beta}}^R\lambda_\alpha \lambda_\beta}\\
\label{eq:F2tr}=&\frac{1}{R} + \sqrt{ \frac{R-1}{R} \left( 1- \tr{\rho_\mathrm{E}^2}\right) },
\end{align}
where we used $\sum_{\alpha\neq\beta}\lambda_\alpha\lambda_\beta=1-\tr{\rho_\mathrm{E}^2}$ in the last step. By plugging Eq.~\eqref{eq:1-tr} into~\eqref{eq:F2tr} and rearranging accordingly, we arrive at
\begin{align*}
\frac{R}{R-1}\left(F^2(\rho_\mathrm{E}^{\mathrm{D}},\rho_\mathrm{E})-\frac{1}{R} \right)\leq \sqrt{1-\mathcal{W}_\mathrm{P}^2}.
\end{align*}
Now, by Eq.~\eqref{eq:Dc}, we can identify the left hand side with $\mathcal{D}_\mathrm{F}$, such that
\begin{align}\label{eq:Connectapp}
\mathcal{D}_\mathrm{F}^2+\mathcal{W}_\mathrm{P}^2 \leq 1.
\end{align}
In consideration of the hierarchy~\eqref{eq:hierarchy}, this proofs Eq.~\eqref{eq:Connect} for the distinguishability measure $\mathcal{D}_\mathrm{F}$, and, thus, finishes the proof of Eq.~\eqref{eq:Connect}. Note that this proof can also be performed using Theorem~1 in Ref.~\cite{Miszczak-SS-2009}. However, we provided the entire proof in order to see that the inequality~\eqref{eq:Connectapp} saturates if and only if $\rho_\mathrm{E}=\rho_\mathrm{E}^{\mathrm{D}}$, given the use of the Cauchy-Schwarz inequality in Eq.~\eqref{eq:CSuse}.

\end{appendix}



%

\end{document}